\documentclass[12pt]{iopart}
\usepackage{graphicx}
\usepackage{subfig}

\begin{document}

\title[Kinetic study of EBW plasma current start-up]{Kinetic study of plasma current start-up under EBW power in tokamak plasmas}



\author{E.J. du Toit$^{1,2}$, M.R. O'Brien$^2$ and R.G.L. Vann$^1$}
\address{$^1$ York Plasma Institute, Department of Physics, University of York, York, YO10 5DD, UK}
\address{$^2$ Culham Centre for Fusion Energy, Abingdon, OX14 3DB, UK}

\ead{ejdt500@york.ac.uk}

\date{}                     



\begin{abstract}
Tokamak plasma current start-up assisted by Electron Bernstein waves (EBW) has been demonstrated successfully in a number of experiments. The dynamic start-up phase involves a change in field topology, as the initially open magnetic field lines form closed flux surfaces (CFS) under the initiation of a plasma current. This change in field topology will bring about a change in the current drive (CD) mechanism, and, although various mechanisms have been proposed to explain the formation of CFS, no detailed theoretical studies have previously been undertaken. Here, we report on the development of a start-up model for EBW-assisted plasma current start-up in MAST. It is shown that collisions are  responsible for only a small part of the CD, while the open magnetic field line configuration leads to an asymmetric confinement of electrons, which is responsible for the greater part of the CD.
\end{abstract}


\section{Introduction}
Non-inductive plasma current start-up is a very important area of research for the spherical tokamak (ST) due to a lack of space in a reactor for a neutron-shielded inboard solenoid. A possible start-up technique, based on the use of radiofrequency (RF) waves, has been proposed and developed in order to avoid a central solenoid in future ST devices \cite{Forest_1994, Ejiri_2006, Uchida_2004, Shevchenko_2007}. The electron Bernstein wave (EBW) start-up technique has proven particularly successful \cite{Shevchenko_2010, Maekawa_2005}, with currents up to $73 \, \textrm{kA}$ achieved noninductively on MAST with up to $100 \, \textrm{kW}$ of input power \cite{Shevchenko_2015}.

\begin{figure}[!hbt]
	\centering
		\hspace*{\fill}
		\subfloat[]{%
			\includegraphics[width=0.3\textwidth]{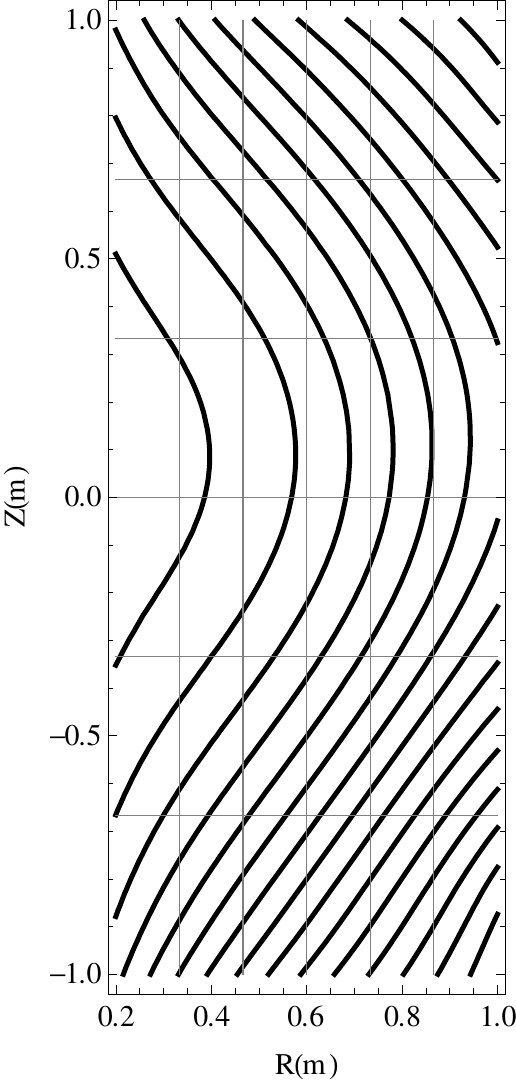}}
		\hfill
		\subfloat[]{%
			\includegraphics[width=0.3\textwidth]{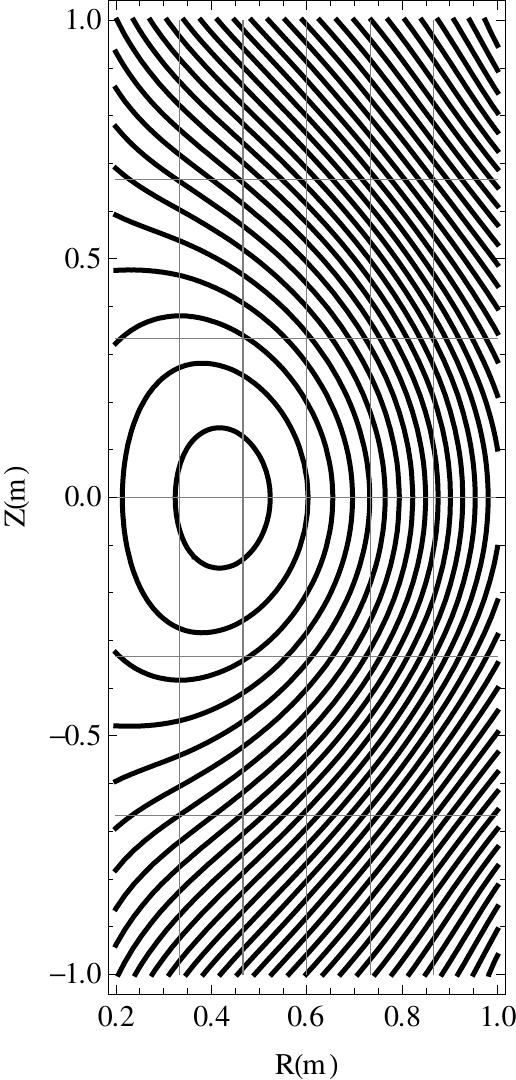}}
		\hspace*{\fill}
	\caption[]{Magnetic flux contours in the poloidal plane for (a) early times, when the field lines are open, and (b) when there is a sufficiently strong plasma current for closed flux surfaces to form. The formation of closed flux surfaces drastically improves plasma confinement, as electrons and ions can otherwise freely stream along the open magnetic field lines out of the plasma volume.}
	\label{fig:field}
	\end{figure}

Among the various phases of noninductive current drive, one of the most important, and most dynamic, is the start-up phase. This phase involves the change of field topology from the open magnetic field line configuration to the formation of closed flux surfaces (CFS), as shown in figure \ref{fig:field}, which has been observed in a number of RF assisted start-up experiments \cite{Forest_1994, Ejiri_2006, Shevchenko_2010, Maekawa_2005}. The formation of CFS drastically affects the plasma equilibrium and confinement, and therefore also the current drive (CD) mechanism. An investigation of the start-up process and development of reliable start-up models are therefore not only important for gaining an understanding of successful start-up in tokamaks, but also for predicting performance and start-up requirements for present and future STs.

The Fisch-Boozer mechanism \cite{Fisch_1980}, based on the preferential heating of electrons with a fixed parallel velocity to produce an anisotropic plasma resistivity, is an attractive concept for CD using electron cyclotron (EC) waves. For typical plasma parameters, however, the core is inaccessible for conventional electromagnetic modes in the range of frequencies corresponding to the first few EC resonances \cite{Shevchenko_2007}. EBWs, however, have been shown to provide localised, highly efficient heating and CD in STs both theoretically \cite{Forest_2000} and experimentally \cite{Shevchenko_2010, Maekawa_2005}. The advantage is that EBWs do not have any density cut-offs inside the plasma, and they are strongly absorbed at nearly all harmonics of the cyclotron resonance, even in relatively cold plasmas \cite{Shevchenko_2007, Shevchenko_2002}.

As EBWs are predominantly electrostatic waves, start-up by this means rely on the mode conversion (MC) from externally launched electromagnetic waves, which can occur through a number of methods \cite{Shevchenko_2010, Maekawa_2005, Petrillo_1987}. The excited EBW is then totally absorbed, and has been observed to generate significant plasma currents \cite{Shevchenko_2010, Maekawa_2005, Shevchenko_2015}.

Experiments concluded that the generated plasma current is carried by energetic electrons, which raises the issue of whether a collision-driven CD mechanism, such as the Fisch-Boozer mechanism, is valid \cite{Maekawa_2005, Yoshinaga_2006}. Alternative CD mechanisms have since been proposed to describe the generation of plasma current and the spontaneous formation of CFS, including a pressure driven current \cite{Ejiri_2006, Maekawa_2005, Yoshinaga_2007, Maekawa_2012}, and the preferential confinement of fast electrons \cite{Maekawa_2012, Ejiri_2007, Yoshinaga_2006}.

Theoretical studies of EBW start-up have mainly relied on equilibrium reconstruction and the study of single particle orbits as a way of describing the initiation of CFS. As such studies are unable to describe the time evolution of observables, they are unable to account for certain experimentally observed effects. In particular, experiments found the following:
	
	\begin{enumerate}

	\item shifting the plasma up or down helps the formation of CFS \cite{Shevchenko_2010, Shevchenko_2015},

	\item larger plasma currents are generated by increasing the vacuum magnetic field strength \cite{Shevchenko_2010, Maekawa_2005, Shevchenko_2015},

	\item and there exists a linear dependence between injected RF power and generated plasma current \cite{Shevchenko_2015}.

	\end{enumerate}

In order to explain these phenomena and show what CD mechanism is responsible for the generated plasma current, a start-up model has been developed in order to interpret and predict start-up performance in current and future STs. This paper describes the development of a kinetic model for studying EBW start-up. It leads to interpretations and predictions with regards to CD mechanisms, plasma-wave interactions, and the role of the vacuum magnetic field.

The paper is structured as follows: the kinetic model is introduced, followed by a discussion of the role of the magnetic field on single particle orbits and electron confinement. Simulations are performed to investigate the CD mechanism and to illustrate the effect of the vacuum magnetic field, before comparisons with experiments are made.

\section{Development of a kinetic model for studying EBW start-up}
The development of start-up models is important to interpret and predict start-up current drive in STs, and provide an understanding of conclusions drawn from experiments. The main components during EBW start-up is expected to be the plasma-wave interaction and the effect of the open magnetic field lines on particle orbits. We have therefore developed a start-up model to study the time evolution of the electron distribution function only, as ions are not expected to dominate the plasma behaviour.  In general, the electron distribution function depends on space, momentum, and time, $f(\vec{r},\vec{p},t)$, and can be used to calculate, amongst other observables, the parallel current density,
	\begin{equation*}
	J(\vec{r},t) = e \int \textrm{d}^3 p \, v_\parallel \, f(\vec{r},\vec{p},t)
	\end{equation*}
and the electron density,
	\begin{equation*}
	n_e(\vec{r},t) = \int \textrm{d}^3 p \, f(\vec{r},\vec{p},t).
	\end{equation*}

The full $6+1$ dimensional distribution function $f(\vec{r},\vec{p},t)$ is a complicated function to study due to the spatial diffusion of electrons - the open magnetic field line configuration means that electrons aren't confined to fixed orbits, but are rather position and momentum dependent - and the plasma-wave interaction being both position and momentum dependent. For this reason, in order to study the time evolution of the distribution function, simplifications are made to ensure the model is tractable and computationally manageable.

The assumption is made that the main physics can be included in a model that is zero dimensional ($0$D) in space, as long as appropriate volume averages are taken when calculating the $0$D approximations. Further, the time it takes for an electron to complete a gyro-orbit is fast compared to all other timescales, such that the momentum dependence can be captured in two dimensions ($2$V) only. The time evolution of the distribution function is then studied in the presence of several effects thought to be important in capturing the main physics during the early stages of the plasma discharge,
	\begin{equation}
	\frac{\partial f}{\partial t} = \textrm{source} - \textrm{loss} + \textrm{EBW heating} + \textrm{collisions} + \textrm{induction}
	\end{equation}
where $f = f(p_\parallel,p_\perp,t)$, and $p_\parallel$ is the momentum along the magnetic field and $p_\perp$ the momentum perpendicular to the magnetic field. In order to include relativistic effects in the EBW heating term, momentum $p$ is used rather than velocity $v$. Additional terms, such as radiative losses or recombination effects, are neglected as these are not expected to dominate the plasma behaviour during start-up.

\subsection{Electron sources}
The source term models cold electrons entering the system, mainly through ionizations. These electrons are assumed to be isotropic in momentum, such that,
	\begin{equation}
	\left( \frac{\partial f}{\partial t} \right)_\textrm{source} = \frac{S_0}{\pi^{3/2} p_0^3} \exp{ \left(- \frac{p_\parallel^2 + p_\perp^2}{p_0^2} \right) }
	\end{equation}
where $p_0$ is the characteristic momentum of these cold electrons, and $S_0$ is the source injection rate.

The value of $S_0$ is chosen to ensure that the density obtained from the distribution function equals some pre-determined density, obtained, for example, from experiments. The time evolution of the density depends on many factors including neutral gas puffing, ionization, recombination, and refuelling from the plasma edge. Models simulating start-up in conventional tokamaks are able to simulate the time evolution of the density \cite{Kim_2012}, but in this paper we will assume the time evolution of the density is known.

\subsection{Collisions}
The collision operator approximates electron-electron collisions, under the assumption that the distribution collides with a background Maxwellian distribution of the same temperature and density, and pitch-angle scattering is due to electron-ion collisions \cite{Karney_1986}.

\subsection{Plasma induction}
In the absence of an external electric field, a change in plasma current will create a loop voltage opposing the change in current according to Lenz's law,
	\begin{equation}
	\label{eq:Lenz}
	V_L = -L_p \frac{dI_P}{dt}
	\end{equation}
where the self-inductance $L_p$ is a function of the plasma major radius $R$, minor radius $a$ and internal inductance $\ell_i$ \cite{Kim_2012, Lloyd_1996},
	\begin{equation*}
	L_p = \mu_0 R \left( \ln \frac{8R}{a} \frac{\ell_i}{2} - 2 \right)
	\end{equation*}
with the internal inductance $\ell_i$ calculated with \cite{Wesson_2004},
	\begin{equation*}
	\ell_i = \frac{2 \int_0^a B_\theta^2 \, r \, \textrm{d} r}{a^2 B_{\theta a}^2}
	\end{equation*}

In the case of a flat $I_P$ profile $\ell_i = 0.5$. This leads to a value for the self-inductance of $L_p = 6.5 \times 10^{-7} \, \textrm{H}$ when using typical MAST parameters.

The presence of an electric field will accelerate electrons along it. In STs, the toroidal magnetic field is about two orders of magnitude stronger than the poloidal field in the vicinity of the magnetic axis during start-up, such that the parallel velocity of electrons is essentially in the toroidal direction. The electric field induced by a change in the toroidal current will therefore affect the parallel motion of electrons,
	\begin{equation}
	\left( \frac{\partial f}{\partial t} \right)_\textrm{induction} = -e \frac{V_L}{2\pi R_0} \frac{\partial f}{\partial p_\parallel}
	\end{equation}
where the value of $V_L$ is obtained under the condition that (\ref{eq:Lenz}) holds and $R_0$ is the major radius. The assumption is made that the majority of electrons will be located around the major radius, hence we take the value of the electric field to be determined at this distance within our $0$D model.

\subsection{EBW heating}
The interaction between the injected RF beam and the plasma is described by an EBW heating term. The EBW has an electric field that is nearly perpendicular to the magnetic field, such that it mainly increases the perpendicular momentum of electrons \cite{OBrien_1986}, and can be described using
	\begin{equation}
	\left( \frac{\partial f}{\partial t} \right)_\textrm{EBW} = \frac{1}{p_\perp} \frac{\partial}{\partial p_\perp} D_0 \left\langle \exp{ \left[ - \left( \frac{\omega - k_\parallel v_\parallel - n \omega_c}{\Delta \omega} \right)^2 \right]} \right\rangle_\textrm{vol} p_\perp \frac{\partial f}{\partial p_\perp}
	\end{equation}
where $D_0$ is a constant to be determined via energy balance (see below), $\omega_c$ is the cyclotron frequency, and a volume average is taken to account for the spatial dependence of absorption. The location of absorption is determined by the resonance condition,
	\begin{equation}
	\label{eq:resonance}
	\omega - k_\parallel v_\parallel - n \omega_c = 0
	\end{equation}
where $\omega_c$ is the relativistic electron cyclotron frequency (ECR). Absorption occurs around the ECR, and is broadened by a relativistic mass shift and a Doppler shift. 

The absorption width $\Delta \omega$ can be related to the resonance width $\Delta R_0$,
	\begin{equation}
	\label{eq:EBW:dR}
	\Delta \omega = \omega \frac{\Delta R_0}{R_0} 
	\end{equation}
where $R_0$ is the radial distance where absorption takes place, or the change in $N_\parallel$, the value of the refractive index parallel to the magnetic field,
	\begin{equation}
	\label{eq:EBW:dN}
	\Delta \omega = \omega \frac{p_\parallel}{m_e c} \Delta N_\parallel
	\end{equation}
depending on the evolution of the EBW as it is absorbed. Typically, the value of $N_\parallel$ evolves along the trajectory of the EBW as it is absorbed, indicating the use of equation (\ref{eq:EBW:dN}). If, instead, we want to highlight the effect of $N_\parallel$ on the generated current, equation (\ref{eq:EBW:dR}) is a better choice for approximating the absorption width, and both equations could be used.

The value of $D_0$ is found by ensuring the correct power is absorbed,
	\begin{equation*}
	P_d = \frac{1}{2} m_e \int \textrm{d} V \, \int v^2 \left( \frac{\partial f}{\partial t} \right)_{\textrm{EBW}} \textrm{d}^3 p
	\end{equation*}
with
	\begin{equation}
	P_d = A \, P_0
	\end{equation}
where $P_0$ is the injected power and the absorption coefficient $A$ is related to the optical depth $\tau$ in the usual way \cite{Petrillo_1987}, $A = 1 - e^{-\tau}$, and is very close to $100 \%$ for a wide range of densities, even in cold plasmas.

It is known that $k_\parallel$ has different signs above and below the midplane, such that EBW rays propagating close to the midplane do not contribute significantly to the current drive as the value of $N_\parallel$ oscillates around zero \cite{Forest_2000, Urban_2011}. If the absorption is localised predominantly above or below the midplane, to gain a directionality with respect to the magnetic field, significant current can be generated \cite{Shevchenko_2002}.

The value of the wavevector parallel to the magnetic field, $k_\parallel$, determines the location of heating for the electrons. The EBW is fully absorbed before it reaches the ECR, such that $\omega - n \omega_c > 0$. If $k_\parallel > 0$, the resonance condition (\ref{eq:resonance}) is satisfied for electrons with $p_\parallel > 0$, and these electrons are heated, while if $k_\parallel < 0$, electrons with $p_\parallel < 0$ are heated. The sign of $k_\parallel$ is determined by the local magnetic field in the region of absorption, as the EBW is essentially perpendicular to the magnetic field.

\section{Orbital losses}
\label{sec:loss}
An important part of start-up is the transition from an open magnetic field line configuration to the formation of CFS. The open magnetic field line configuration allows electrons to freely stream out of the plasma volume. This loss mechanism is charaterized by a loss time $\tau_\textrm{loss}$,
	\begin{equation*}
	\frac{1}{\tau_\textrm{loss}} = \frac{1}{\tau_\parallel} + \frac{1}{\tau_\perp}
	\end{equation*}
where the parallel loss time, due to electrons streaming along field lines, is modelled by the expression
	\begin{equation}
	\tau_\parallel = a \, \exp{\left( \frac{I_\textrm{\small{P}}}{I_\textrm{\small{CFS}}} \right)} \frac{B_\phi}{B_Z} \bigg/ v_\parallel
	\end{equation}
with $a$ the minor radius, $I_\textrm{\small{CFS}}$ the value of the plasma current where CFS first start to form, and $B_Z$ is the $\hat{Z}$-component of the magnetic field. The exponential factor has been included to correct for the fact that parallel loss times become longer once CFS form \cite{Kim_2012}. 

The perpendicular loss time describes the loss of electrons traveling across magnetic field lines due to collisions. Bohm diffusion is adopted to describe this effect \cite{Wauters_2011},
	\begin{equation}
	\tau_\perp = 8a^2 \frac{B_\phi}{T_e(\textrm{eV})}.
	\end{equation}

The loss time does not take into account the spatial structure of the magnetic field, and therefore electrons with the same velocity, but moving in opposite directions, will be lost at the same rate. Wong \cite{Wong_1980} noticed that, in the case of an open magnetic field line configuration, certain electrons can be confined by adding a small vertical magnetic field. The guiding centre equations, which predict a curvature and $\nabla B$ drift, give the following expression for the motion in the $Z$-direction,
	\begin{equation}
	V_Z = \frac{B_Z}{B} v_\parallel - \frac{m_e}{q_e B R} \left( \frac{v_\perp^2}{2} + v_\parallel^2 \right).
	\end{equation}

For $B_Z > 0$ and $B_\phi > 0$, some electrons with $v_\parallel > 0$ will have $V_Z = 0$, and will not be lost, while all other electrons will be lost to either the top or the bottom of the vessel. The vacuum magnetic field can therefore be used to control which electrons satisfy $V_Z = 0$ and are preferentially confined, leading to the initiation of a current. This phenomenon has been used to describe the initiation of CFS using single particle orbits \cite{Maekawa_2012, Ejiri_2007, Yoshinaga_2006}, and will be used here in order to study the confinement of electrons. 

The preferential confinement of electrons results in a net current, and it is therefore important to model the time evolution of this confinement. The spatial structure of the magnetic field, along with the initial position and velocity, determines whether an electron will be lost or confined, but as we cannot study the orbit of every electron, some simplifications need to be made. We assume that the loss term can be described through the simple form,
	\begin{equation}
	\left( \frac{\partial f}{\partial t} \right)_\textrm{\small{loss}} = - \frac{f}{\tau_\textrm{\small{loss}}} P_\textrm{\small{loss}} (p_\parallel,p_\perp)
	\end{equation}
where $P_\textrm{\small{loss}}(p_\parallel,p_\perp)$ is the probability of an electron being lost or confined.

In order to quantify the evolution of the confinement of electrons $P_\textrm{loss}(p_\parallel,p_\perp)$ as a function of plasma current and vacuum magnetic field, we study single particle orbits under the assumption that energetic electrons originate from the ECR layer where they interact with the injected RF beam. Energetic electrons are most likely to be lost, and can only be created through an interaction with the injected RF beam. As absorption occurs around the ECR layer, we therefore study particle orbits originating from the midplane at the ECR layer. The orbit of an electron is traced out, and if it returns to its starting position, it is considered to be confined (whether it be a trapped or passing orbit). The collection of initial velocities which leads to confined orbits can be calculated, and it is this function that $P_\textrm{\small{loss}}(p_\parallel,p_\perp)$ represents.

From studying electron orbits, and the confinement of electrons originating from different locations, the assumption is made that the probability of an electron being lost or confined can be represented by,
	\begin{equation}
	P_\textrm{\small{loss}}(p_\parallel,p_\perp) = 1 - \exp{ \left( -\ell \frac{p_\parallel^2}{p_\perp^2} \right) }
	\end{equation}
where $\ell = \ell(p_\parallel,I_P,I_\textrm{\small{CFS}},Z_0)$ contains the dependence of the loss term on a number of different factors, including the plasma current $I_P$, the value of the plasma current where CFS first start to form $I_\textrm{CFS}$, and the vertical shift of the plasma $Z_0$. Of course, $P_\textrm{\small{loss}} = 0$ if an electron is confined, with the probability increasing exponentially to $P_\textrm{loss} = 1$ for a lost electron.

This $0$D approach to quantifying the loss term only considers which electrons are lost, and not where they are lost to or what happens with them after they are lost out of the plasma. As ions are not considered either, any charge build-up or electric fields generated by charge seperation are neglected, in order to ensure the model is tractable.

\subsection{Evolution of electron confinement}
\begin{figure}[!hbt]
	\centering
		\subfloat[]{%
			\includegraphics[width=0.25\textwidth]{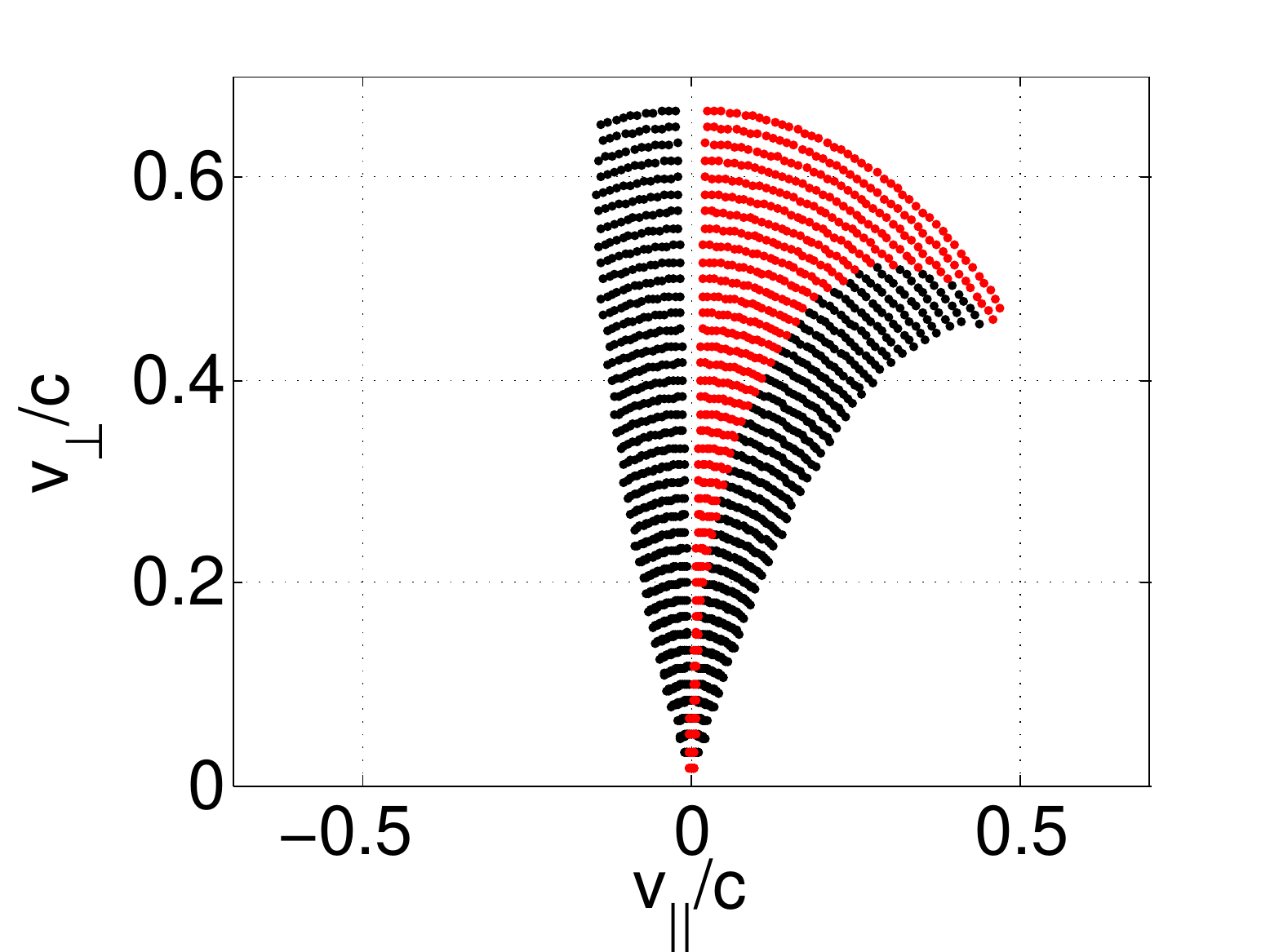}}
		\hfill
		\subfloat[]{%
			\includegraphics[width=0.25\textwidth]{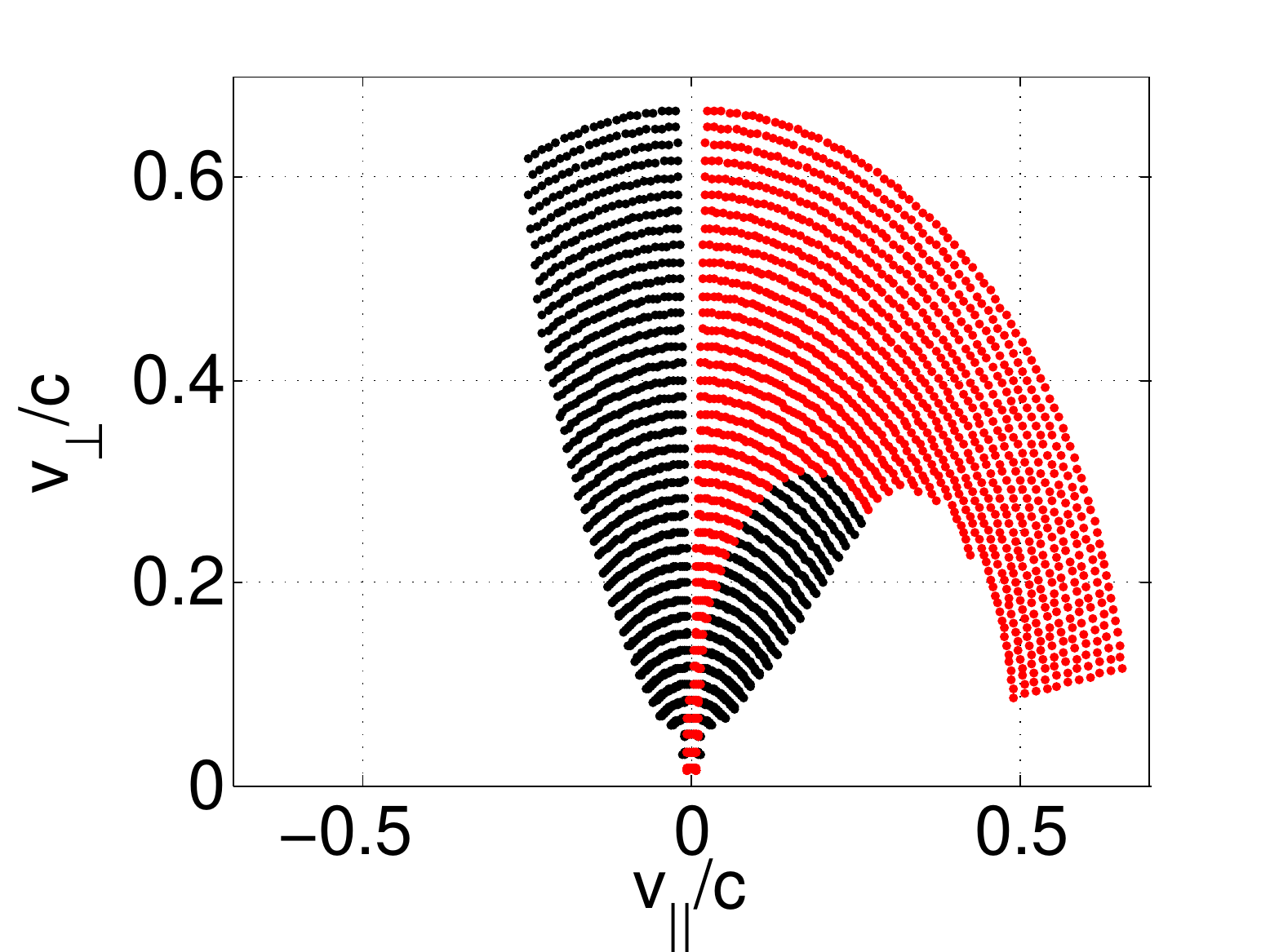}}
		\hfill
		\subfloat[]{%
			\includegraphics[width=0.25\textwidth]{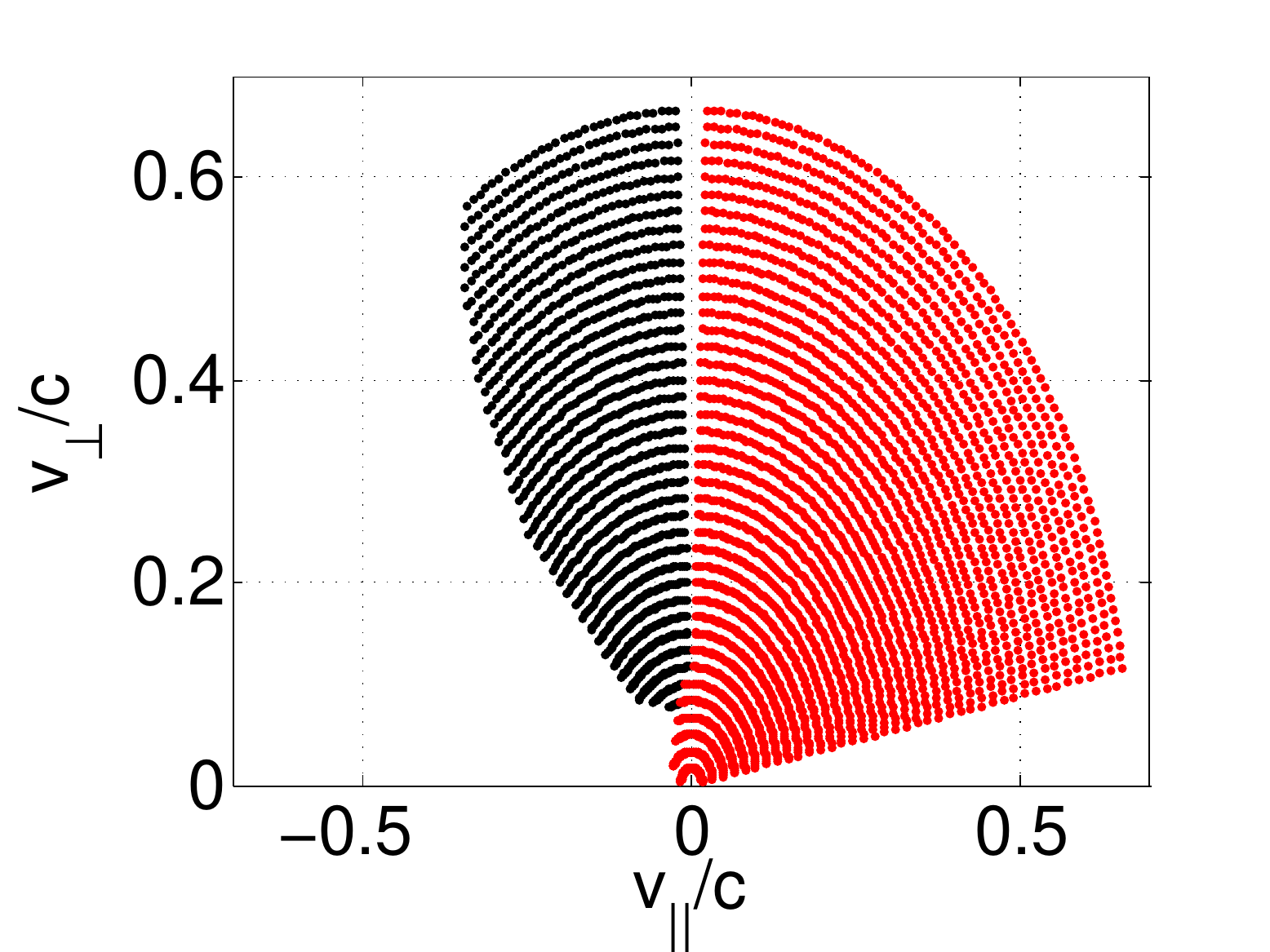}}
		\hfill
		\subfloat[]{%
			\includegraphics[width=0.25\textwidth]{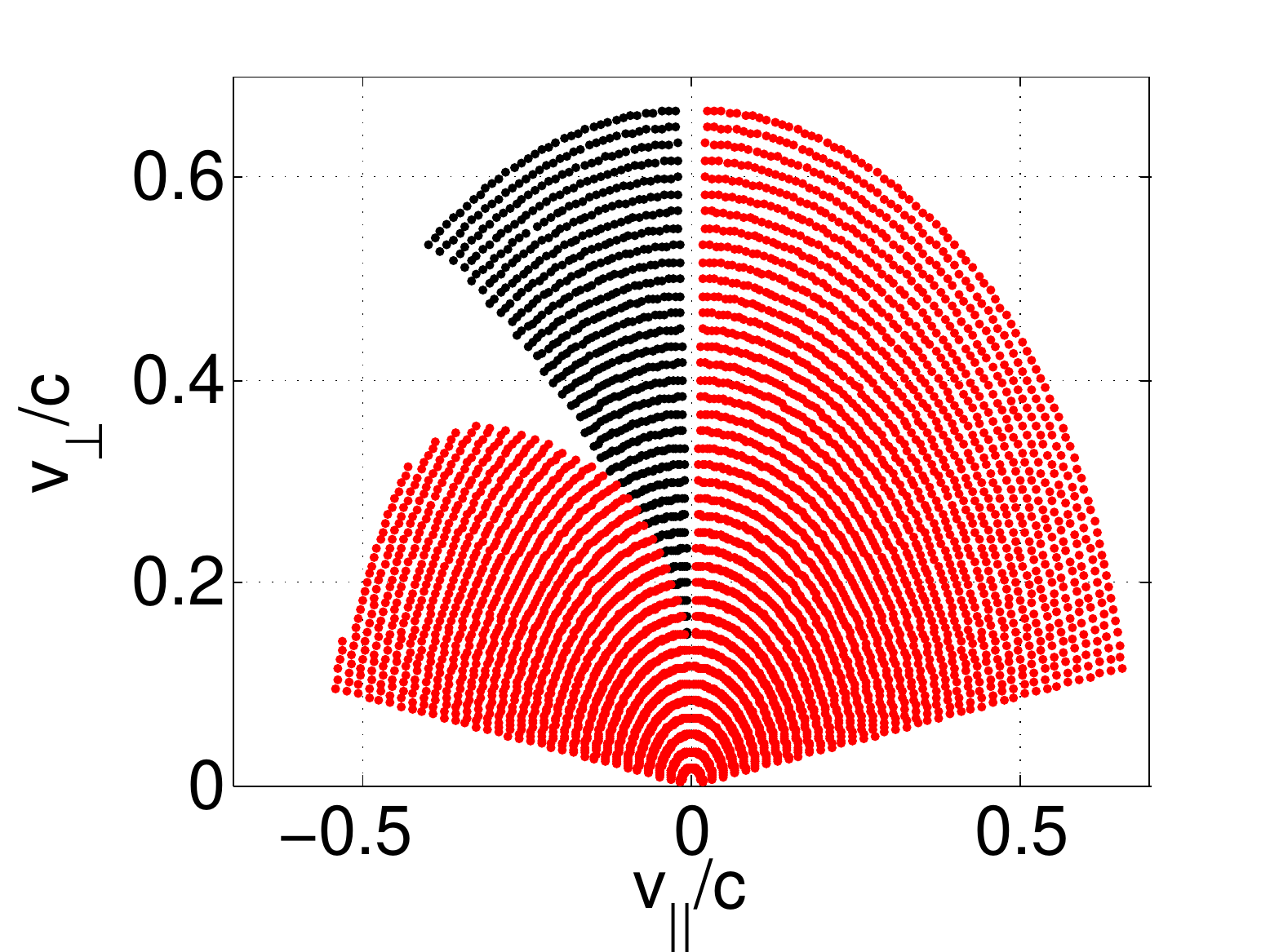}}
	\caption[]{The initial velocities of confined electrons originating from the ECR layer to form passing (red) or trapped (black) orbits for $B_V = 10 \, \textrm{mT}$ and increasing plasma current, (a) $I_P = 5 \, \textrm{kA}$, (b) $I_P = 10 \, \textrm{kA}$, (c) $I_P = 15 \, \textrm{kA}$, and (d) $I_P = 20 \, \textrm{kA}$. The first CFS start to form when all forward electrons, with $v_\parallel > 0$, are confined, at $I_P = I_\textrm{\small{CFS}} = 15 \, \textrm{kA}$ in this case.}
	\label{fig:loss:Ip}
	\end{figure}

The generation of a plasma current creates a self-field $B_a$ which changes the total magnetic field, and therefore also the confinement of electrons. Electrons originating from the midplane, at the ECR layer, are considered confined (trapped or passing) if they return to their starting point, and are plotted in figure \ref{fig:loss:Ip} for increasing plasma current. In this case, the vacuum magnetic field $B_V$ consists of constant, vertical field with $B_V = 10 \, \textrm{mT}$.

Initially, only those electrons that satisfy $V_Z = 0$, and some electrons around it due to the mirror shaped magnetic field, are confined. The asymmetric confinement area expands towards the lower energy region as $I_P$ increases, until all forward electrons are confined. At this point, the self-field equals the vacuum magnetic field, the first CFS start to form, and $I_P = I_\textrm{CFS}$ by definition. Further increasing the current leads to an increase in confinement of electrons moving opposite to the magnetic field, until all electrons are confined and the last CFS encloses a substantial volume.

Even in this simple example it is clear that electrons with $p_\parallel > 0$ and $p_\parallel < 0$ have different confinement, and the optimal operating region will be around $I_P = I_\textrm{CFS}$, where all forward electrons, with $p_\parallel > 0$, are confined.

\subsection{Dependence on $I_\textrm{\small{CFS}}$}
The confinement of electrons does not only depend on the plasma current, but also on the vacuum magnetic field strength and shape, as well as the current density profile. All of these dependences are contained within a single parameter, $I_\textrm{\small{CFS}}$. 

\begin{figure}[!hbt]
	\centering
		\subfloat[]{%
			\includegraphics[width=0.25\textwidth]{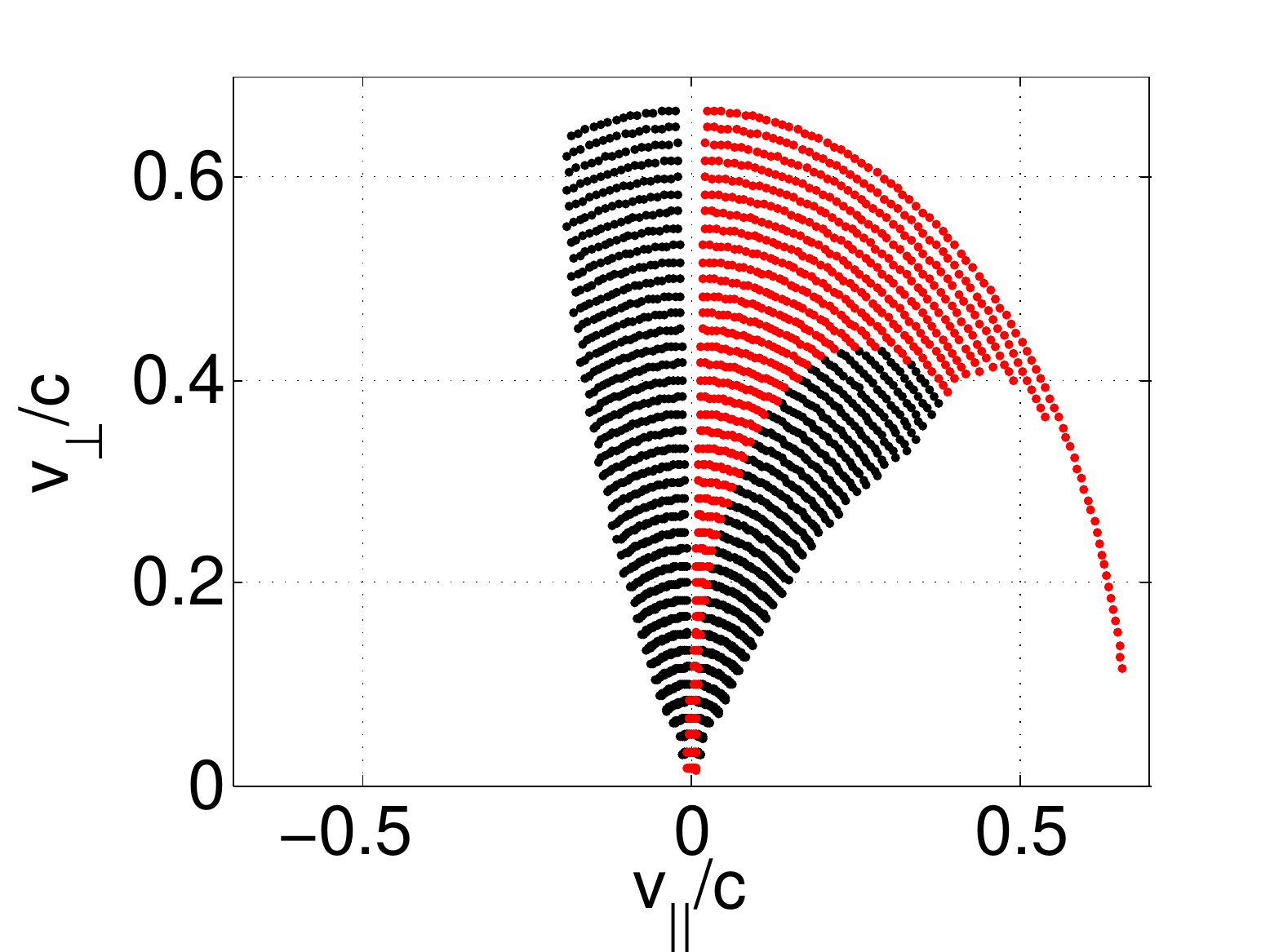}}
		\hfill
		\subfloat[]{%
			\includegraphics[width=0.25\textwidth]{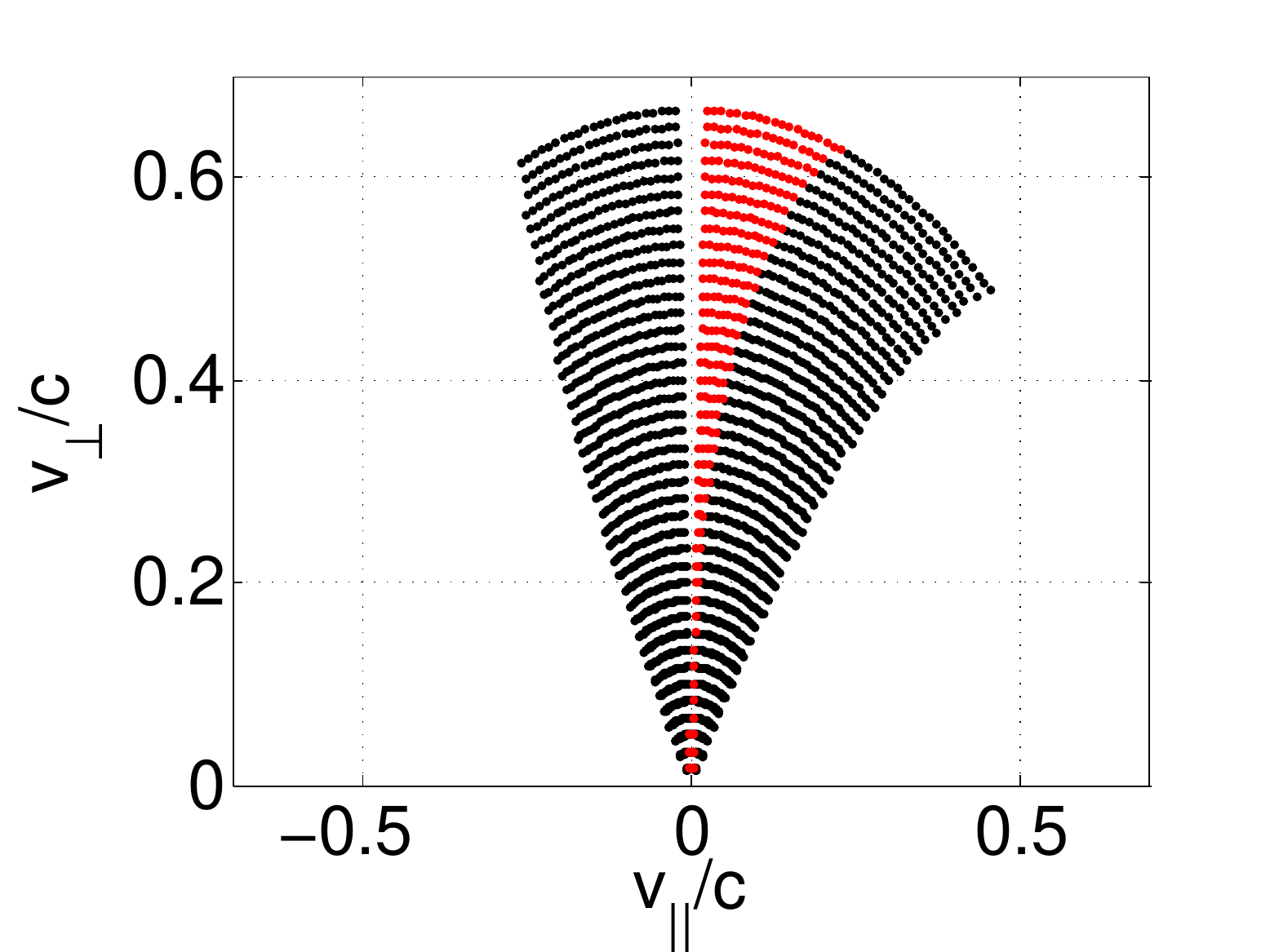}}
		\hfill
		\subfloat[]{%
			\includegraphics[width=0.25\textwidth]{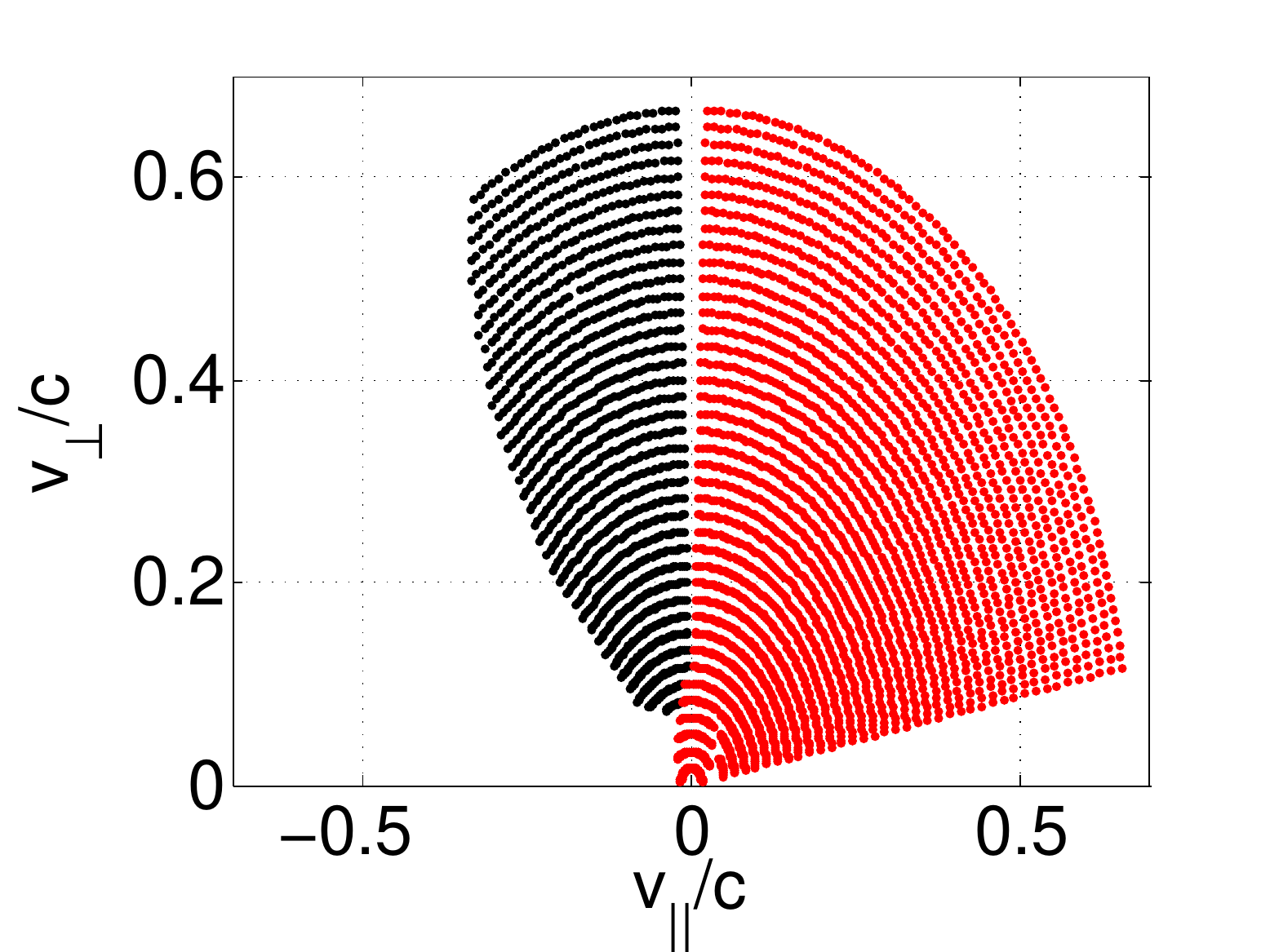}}
		\hfill
		\subfloat[]{%
			\includegraphics[width=0.25\textwidth]{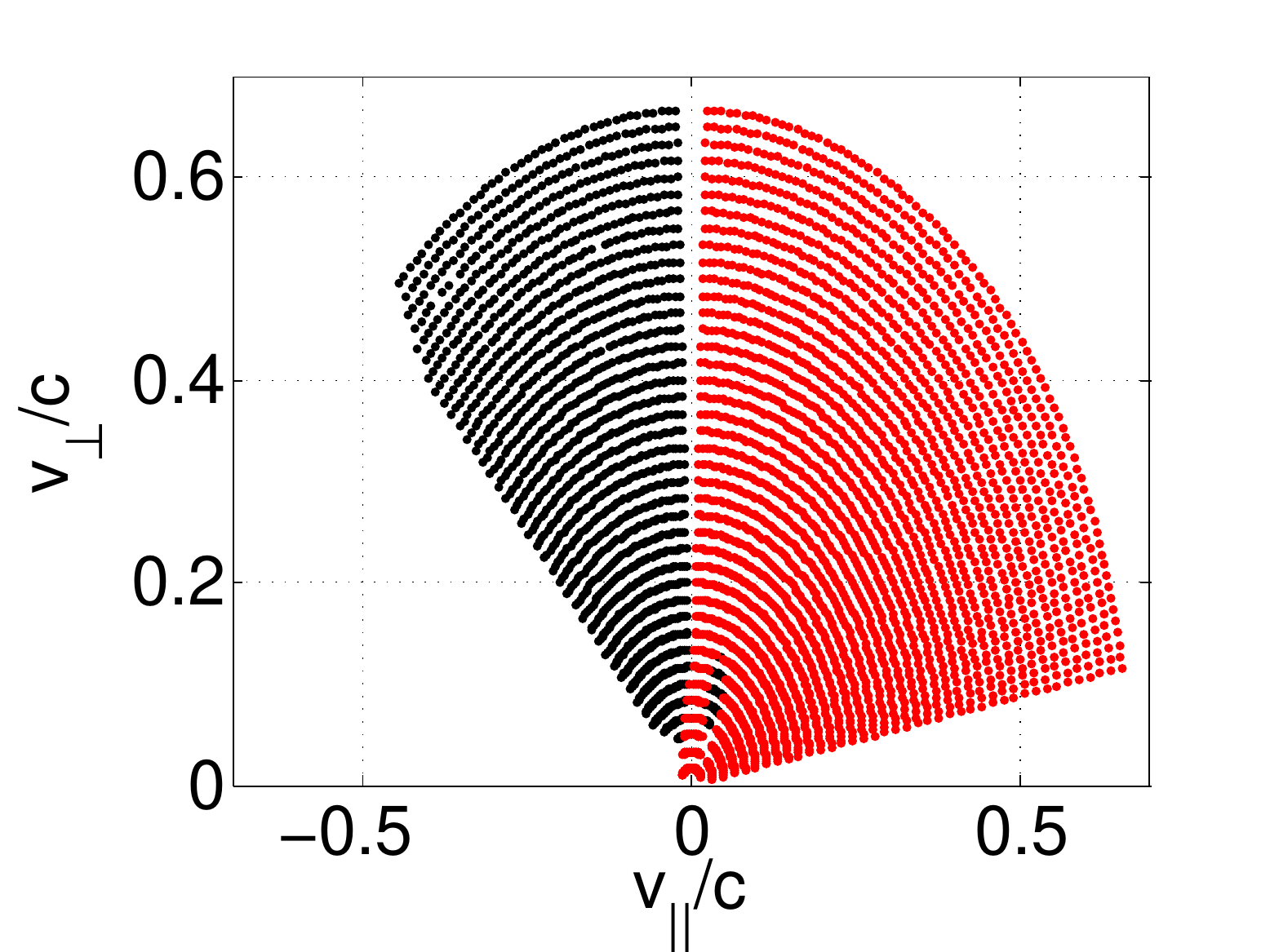}}
	\caption[]{The initial velocities of electrons originating from the ECR layer to form passing (red) or trapped (black) orbits for a constant, vertical vacuum magnetic field and ratio $I_P/I_\textrm{\small{CFS}} = 1/2$ with (a) $B_V = 10 \, \textrm{mT}$ and (b) $B_V = 20 \, \textrm{mT}$, and $I_P/I_\textrm{\small{CFS}} = 1$, with (c)$B_V = 10 \, \textrm{mT}$ and (d)  $B_V = 20 \, \textrm{mT}$. Apart from the confinement of fast electrons with large $p_\parallel$, there is almost no difference in the confinement of electrons when the ratio $I_P/I_\textrm{\small{CFS}}$ is the same.}
	\label{fig:loss:Bv}
	\end{figure}

Figure \ref{fig:loss:Bv} shows the confinement of electrons for a constant, vertical magnetic field of $B_V = 10 \, \textrm{mT}$ and $B_V = 20 \, \textrm{mT}$. The confinement is compared for $I_P = \frac{1}{2} I_\textrm{\small{CFS}}$ and $I_P = I_\textrm{\small{CFS}}$. Remarkably, the only differences are the confinement of fast electrons with large $p_\parallel$, while the confinement of electrons with the same ratio $p_\parallel/p_\perp$ is largely unaffected by the strength of the vacuum magnetic field, as long as the ratio of $I_P/I_\textrm{\small{CFS}}$ is the same.

The value of $I_\textrm{\small{CFS}}$ is related to the strength and shape of the vacuum magnetic field, and the shape of the current density profile. These dependencies only influences the value of $I_\textrm{\small{CFS}}$, and, as long as the ratio of $I_P/I_\textrm{\small{CFS}}$ is the same, the confinement of electrons remains very similar, as illustrated in figure \ref{fig:loss:Bv}.

The complex nature of the spatial dependence of elecron confinement is therefore contained within a single parameter, $I_\textrm{\small{CFS}}$, which is equal to the plasma current at which all forward electrons are confined, and the first CFS start to form. As there could be no certainty with the particular choice of the current density profile, and a different profile will lead to a different value of $I_\textrm{\small{CFS}}$, the value of $I_\textrm{\small{CFS}}$ could easily be adjusted to allow for this uncertainty.

\subsection{Effect of a vertical shift $Z_0$}
\begin{figure}[!hbt]
	\centering
		\subfloat[]{%
			\includegraphics[width=0.25\textwidth]{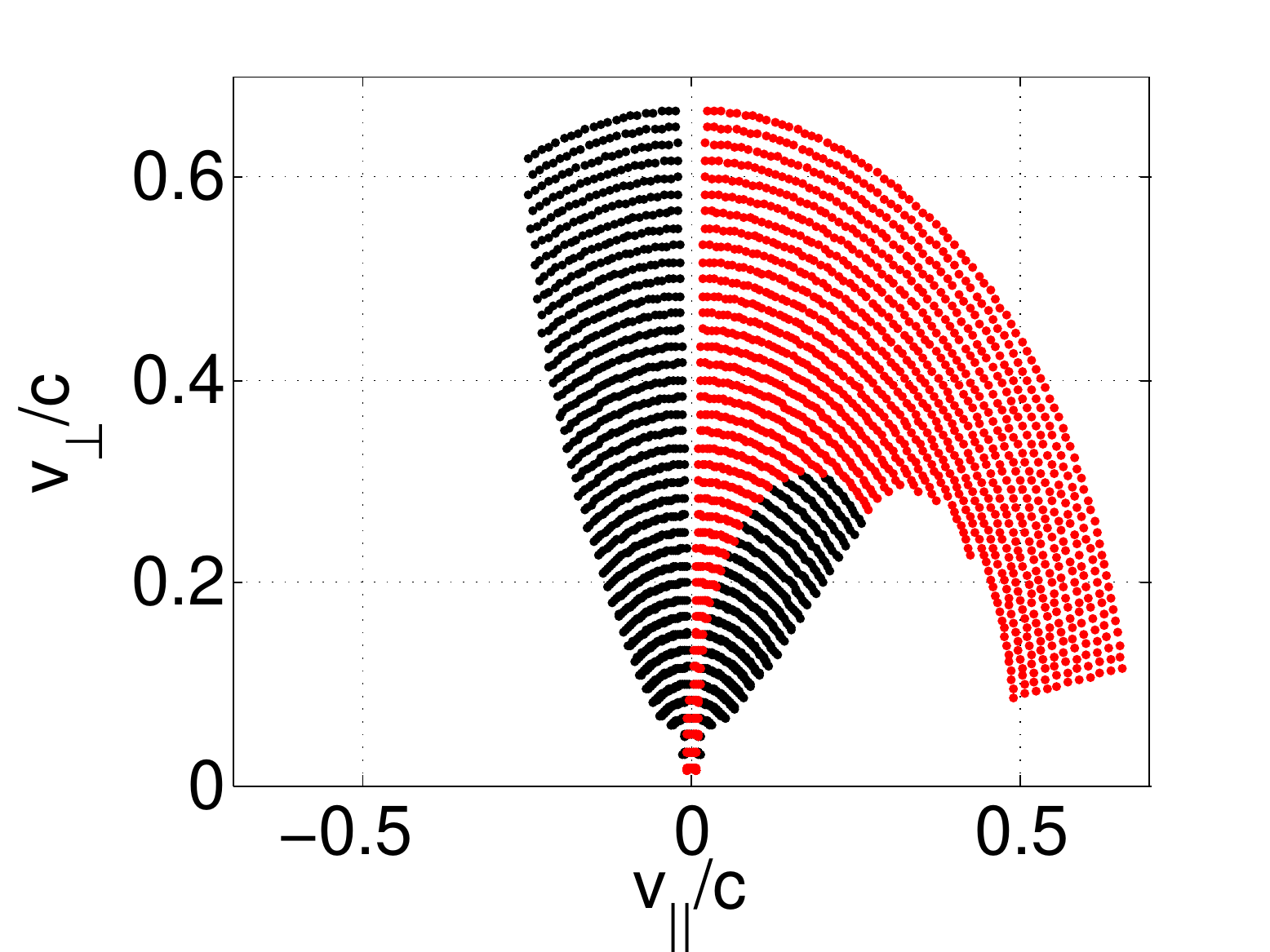}}
		\hfill
		\subfloat[]{%
			\includegraphics[width=0.25\textwidth]{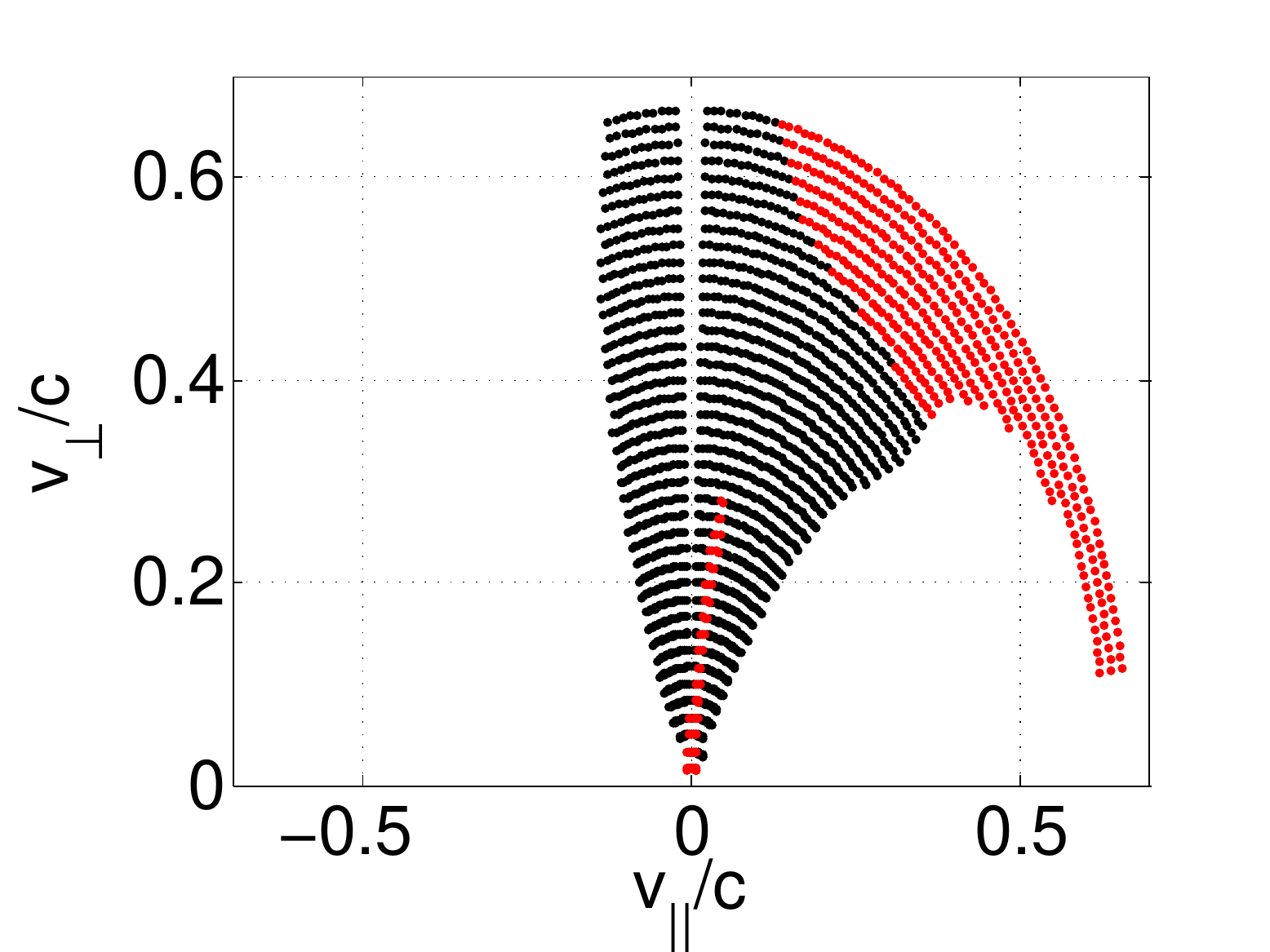}}
		\hfill
		\subfloat[]{%
			\includegraphics[width=0.25\textwidth]{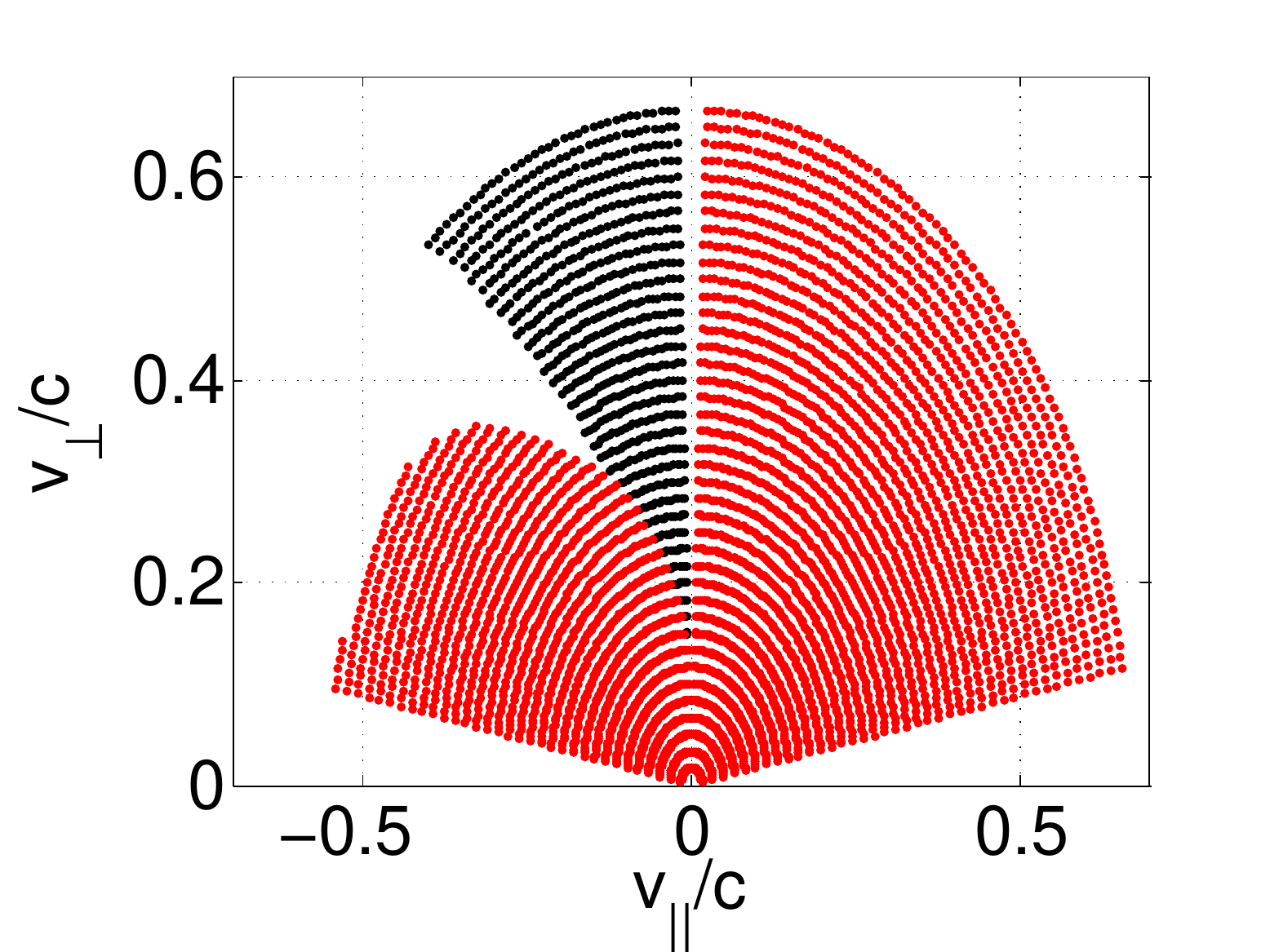}}
		\hfill
		\subfloat[]{%
			\includegraphics[width=0.25\textwidth]{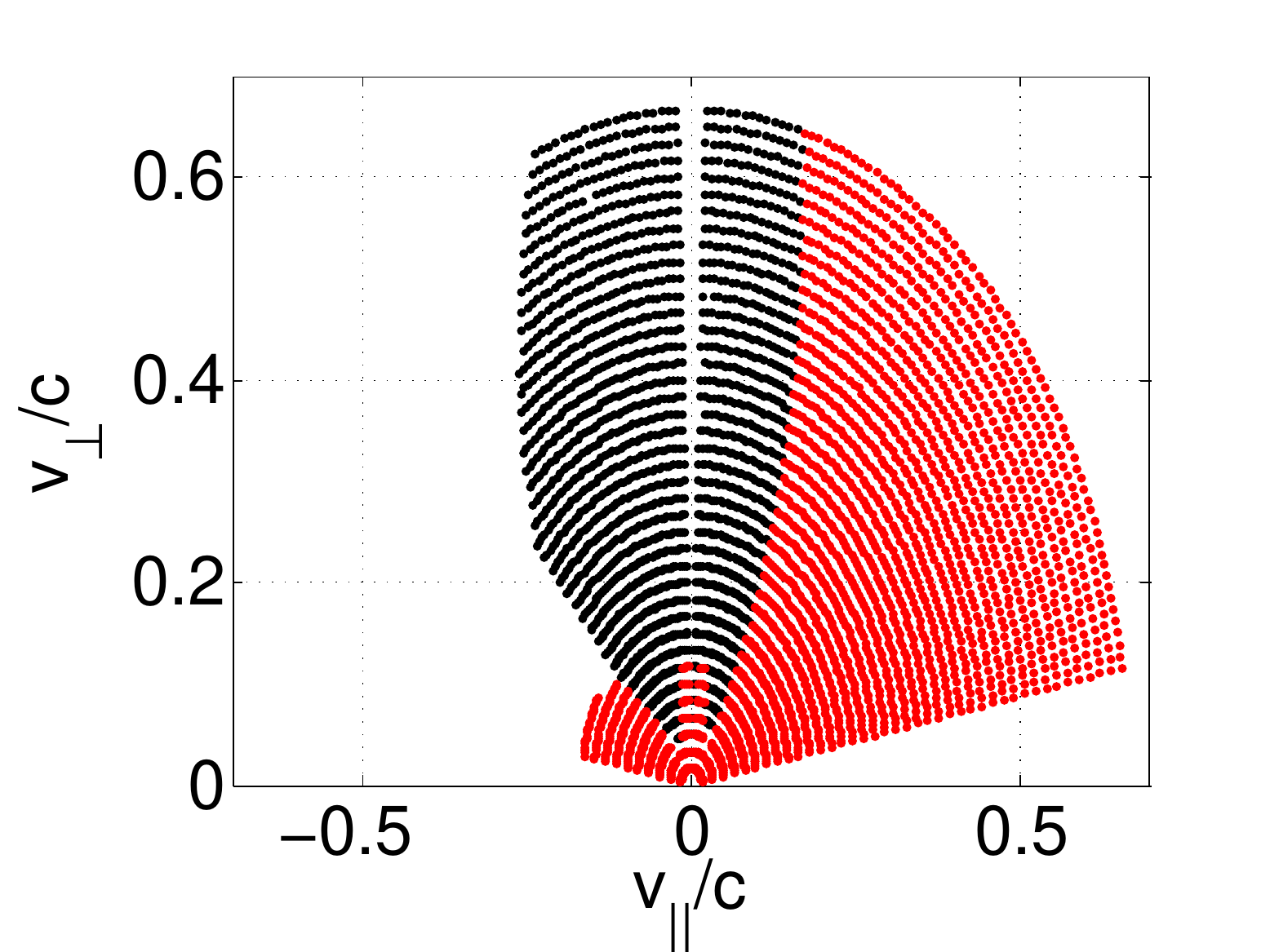}}
	\caption[]{The initial velocities of electrons originating from the ECR layer to form passing (red) or trapped (black) orbits for a constant, vertical vacuum magnetic field of $B_V = 10 \, \textrm{mT}$, with  $I_P = 10 \, \textrm{kA}$ and (a) $Z_0 = 0 \, \textrm{cm}$ and (b) $Z_0 = 40 \, \textrm{cm}$, and $I_P = 20 \, \textrm{kA}$ with (c) $Z_0 = 0 \, \textrm{cm}$ and (d) $Z_0 = 40 \, \textrm{cm}$. Shifting the plasma up or down acts to enhance the asymmetry by reducing the confinement of electrons with $p_\parallel < 0$. The confinement of electrons with large $p_\parallel > 0$ is also affected, but there are very few electrons in this region of momentum space.}
	\label{fig:loss:Z0}
	\end{figure}

Experiments conducted on MAST indicate that shifting the plasma up or down helps to form CFS. A vertical shift is generated by creating a radial field $B_R$, such that the point where $B_Z = 0$ shifts up or down by a distance $Z_0$. The impact such a shift has on the confinement of electrons for a constant, vertical vacuum magnetic field of $B_V = 10 \, \textrm{mT}$, is shown in figure \ref{fig:loss:Z0}. This vertical shift acts to enhance the asymmetry of electron confinement, by reducing the confinement of electrons moving opposite to the magnetic field. The confinement of fast electrons with large $p_\parallel > 0$ is also affected, but there would be very few electrons in this region of momentum space, as the EBWs mainly create fast electrons with large $p_\perp$. The effect of reducing the confinement of these electrons is therefore expected to be minimal.

\newpage
\section{EBW start-up in MAST}
We now turn our attention to modelling microwave start-up in MAST. The EBW start-up method employed here relied on the production of low-density plasma by RF pre-ionization around the fundamental ECR, and a subsequent double mode conversion (MC) for the excitation of EBWs.  Using a $28 \, \textrm{GHz}$ gyrotron, placing the ECR at $R = 0.4 \, \textrm{m}$, capable of delivering $100 \, \textrm{kW}$ for up to $0.5 \, \textrm{s}$, significant plasma current was achieved at an efficiency of approximately $1 \, \textrm{A}/\textrm{W}$ \cite{Shevchenko_2010, Shevchenko_2015}.

The start-up scheme consisted of the MC of an ordinary (O) mode RF wave, incident from the low field side of the tokamak, into the extraordinary (X) mode with the help of a grooved mirror-polarizer incorporated in a graphite tile on the central rod. The X-mode, propagating back into the plasma, passed through the ECR and experienced a subsequent slow X to EBW MC near the upper hybrid resonance (UHR). The excited EBW mode was totally absorbed before it reached the ECR, due to the Doppler shifted resonance.

The EBW is essentially perpendicular to the UHR layer, such that $N_\perp$, the perpendicular component of the refractive index, exceeds $N_\parallel$, the component of the refractive index parallel to the magnetic field, by about two orders of magnitude. EBWs can, however, develop large values for the wave vector, and therefore $N_\parallel$ can reach values $N_\parallel > 1$. In experiments EBWs developed $N_\parallel$ within the range $0.3-0.5$ as they approach the ECR \cite{Shevchenko_2015b}, while EBW with $N_\parallel = 1$ were found to be responsible for the creation of fast electrons \cite{Shevchenko_2010}.

Density measurements indicated that the density was about $\sim 3 \times 10^{17} \, \textrm{m}^{-3}$. It is reasonable to assume that, through ionization, the density will increase during start-up, but its exact time evolution depends on refuelling from the edge, recombination, ionization and other losses, including orbital losses. For simplicity, we assume a fixed density evolution,
	\begin{equation*}
	n_e = n_{e0} \bigg( 0.1 + 0.9 \textrm{tanh}\left[ \frac{t}{t_0} \right] \bigg)
	\end{equation*}	
where $n_{e0} = 3 \times 10^{17} \, \textrm{m}^{-3}$ and $t_0 = 0.05 \, \textrm{s}$. In order to compare different start-up scenarios, we will assume that the density evolution is always the same, regardless of the number of electrons lost. Although this is not physical, experimental measurements of density are not readily available, and this allows us to isolate and study certain effects during start-up that cannot be studied experimentally.

There are three parameters that are fitted during a simulation: the number of electrons entering the system per unit time $S_0$ ensures the correct density evolution; the absorption constant $D_0$ ensures the correct power is absorbed; and the loop voltage $V_L$, resulting from plasma induction, ensures Lenz's law is always satisfied.

\subsection{Collisional start-up}
\begin{figure}[!hbt]
	\centering
		\subfloat[]{%
			\includegraphics[width=0.33\textwidth]{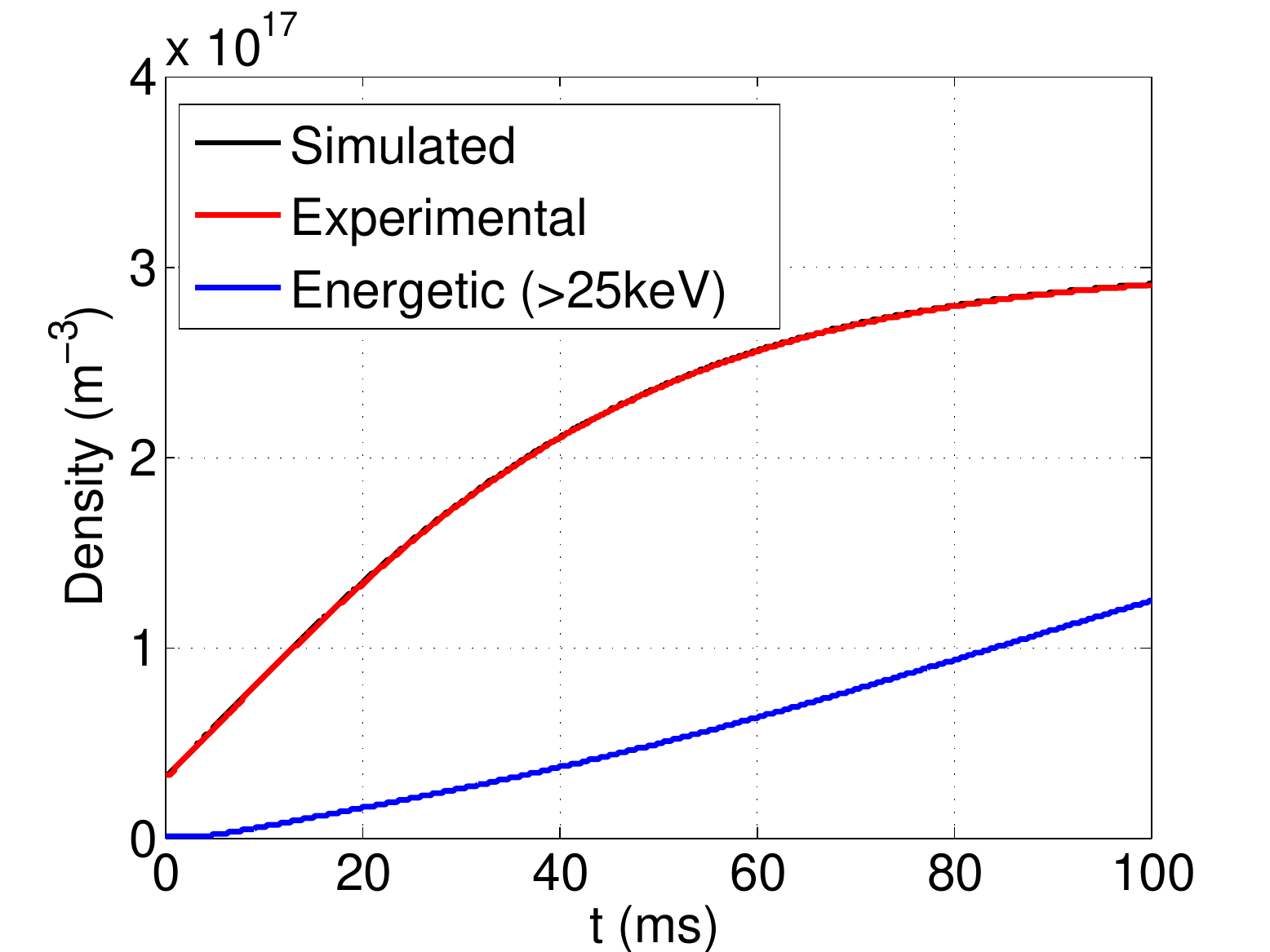}}
		\hfill
		\subfloat[]{%
			\includegraphics[width=0.33\textwidth]{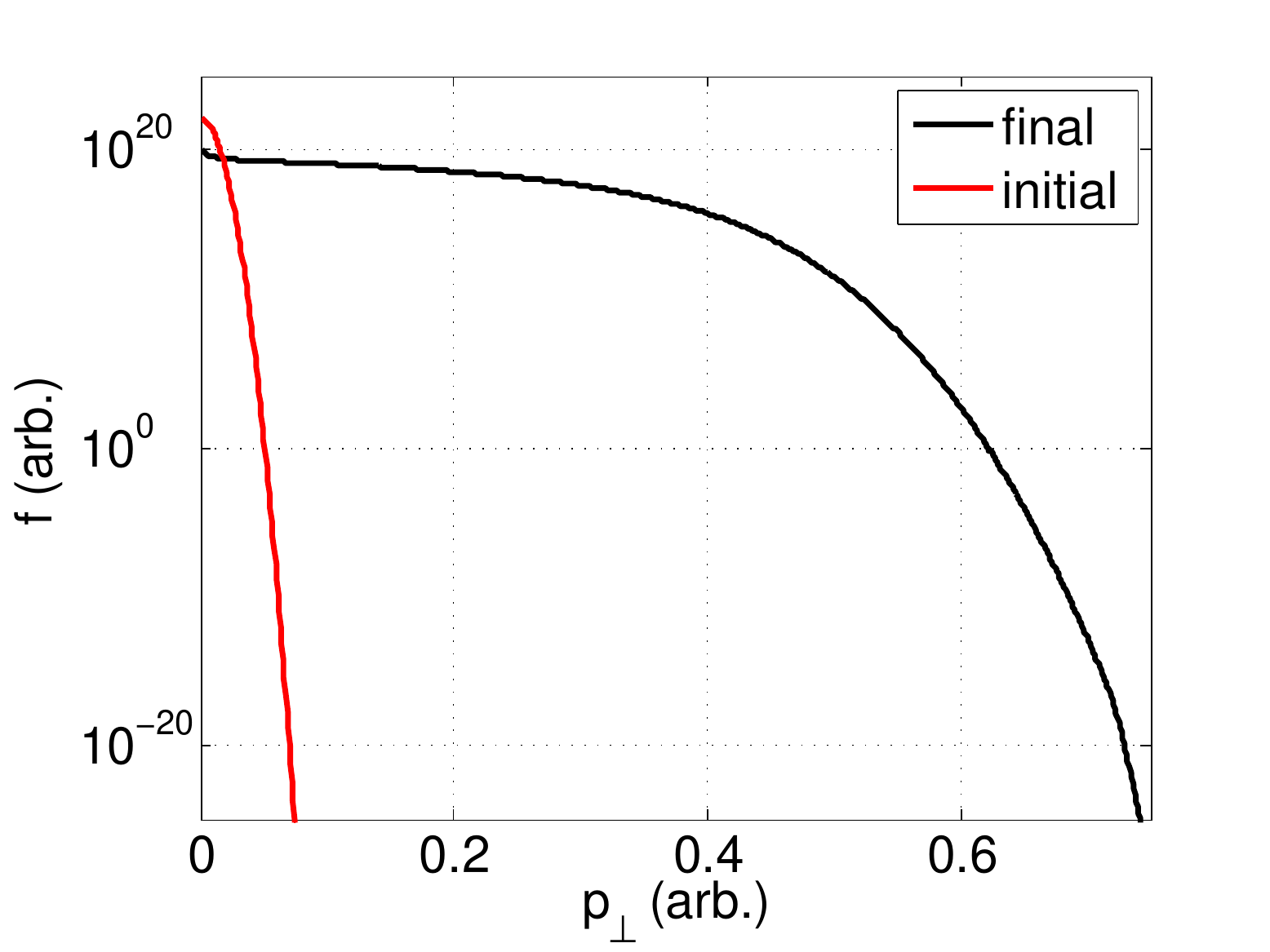}}
		\hfill
		\subfloat[]{%
			\includegraphics[width=0.33\textwidth]{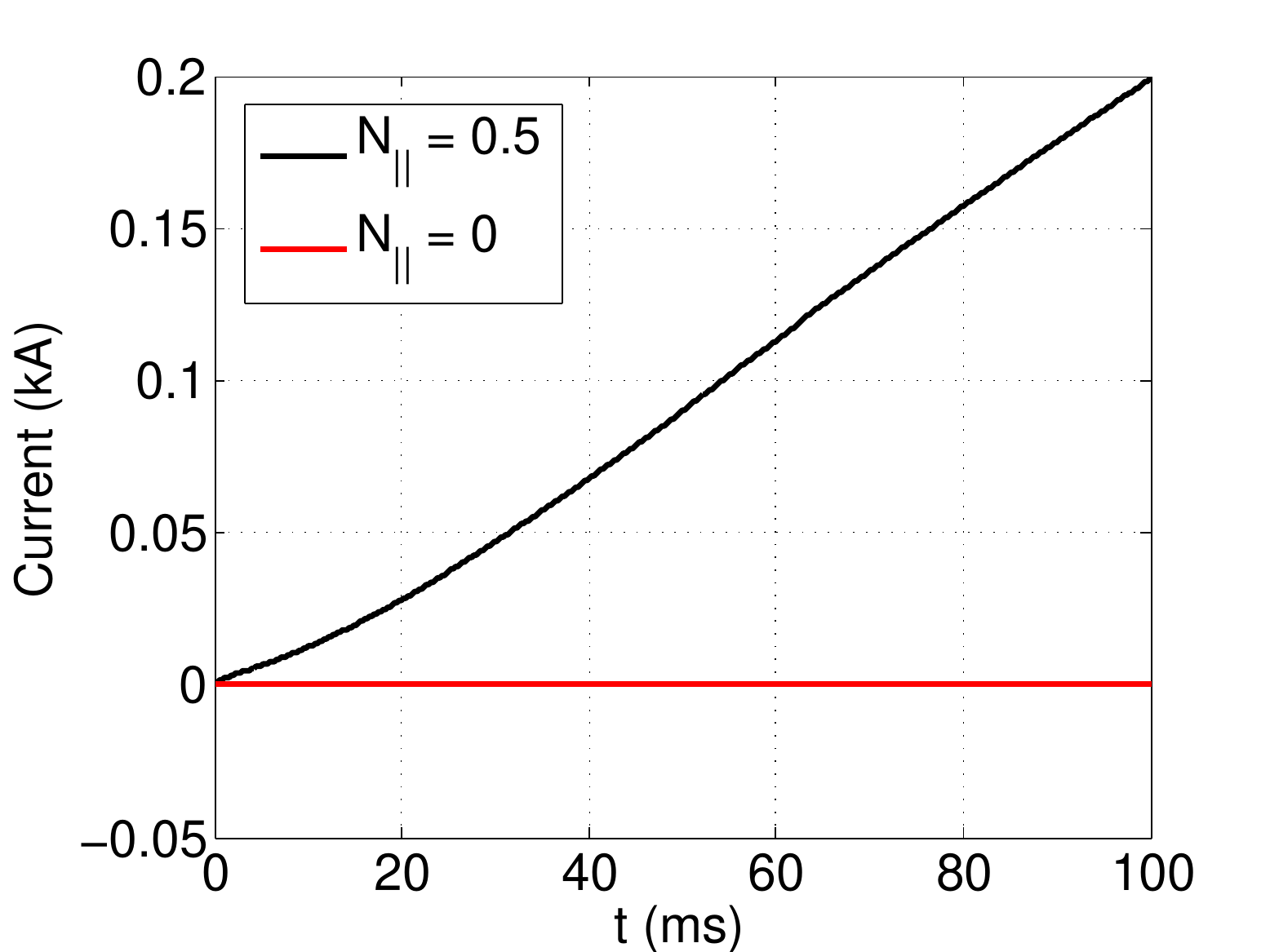}}
	\caption[]{The time evolution of (a) the density shows the simulated and experimental densities being equal, as well as the density of energetic electrons (with energies greater than $25 \, \textrm{keV}$), which are created through the flattening of the distribution function, as shown in (b) - in order for the power to be absorbed, electrons are accelerated to higher energies. The time evolution of (c) the current generation shows that a current can be generated by preferentially heating electrons, similar to the Fisch-Boozer mechanism.}
	\label{fig:startup:collisions}
	\end{figure}

The Fisch-Boozer mechanism, based on the preferential heating of electrons with a fixed parallel velocity to produce an anisotropic plasma resistivity, is an attractive concept for CD using EC waves. This CD mechanism is driven by collisions, and has been all but excluded as the CD mechanism during EBW start-up due to the current being carried by energetic electrons which undergo very few collisions. In order to see whether energetic electrons are formed under EBW power, and if collisions can generate a current, consider a start-up simulation,
	\begin{equation*}
	\frac{\partial f}{\partial t} = \textrm{source} + \textrm{RF heating} + \textrm{collisions} + \textrm{loop voltage}
	\end{equation*}
where all electron losses are excluded. We assume the EBW has a well-defined wave vector, such that $N_\parallel$ is fixed, and $\Delta R = 0.05 \, \textrm{m}$ in (\ref{eq:EBW:dR}). Figure \ref{fig:startup:collisions} compares two cases: $N_\parallel = 0$ and $N_\parallel = 0.5$.

Firstly, it is illustrated, for $N_\parallel = 0.5$, that the simulated density is fitted to the experimental density, and a population of energetic electrons starts to form due to the flattening of the distribution function. In order for all the power to be absorbed, electrons are accelerated to increasingly higher energies, and due to the large value of $N_\parallel$, even fast electrons can still interact with the EBW.

\begin{figure}[!hbt]
	\centering
	\includegraphics[width=1\textwidth]{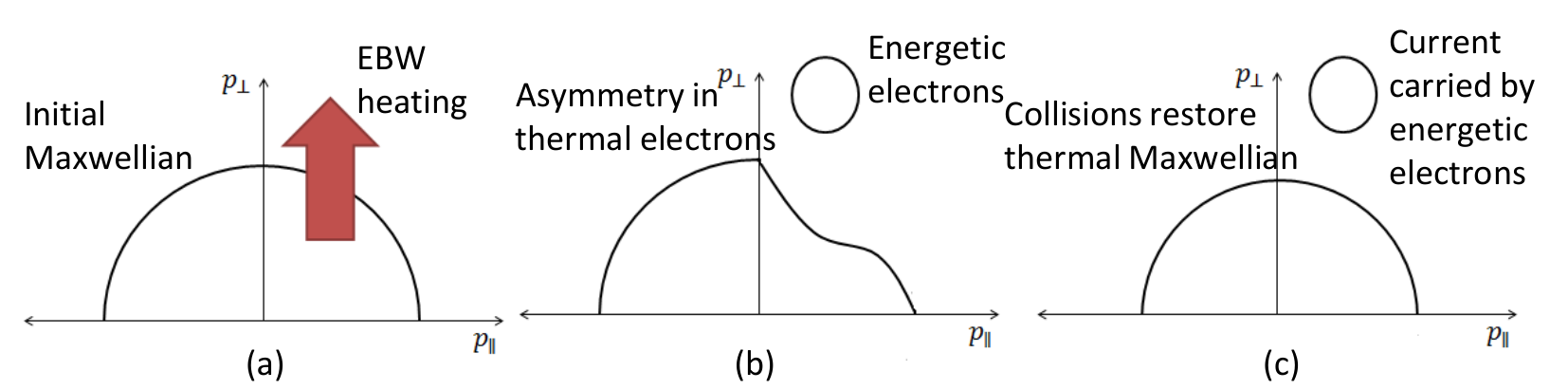}
	\caption[]{Schematic of the collisional current-drive: (a) The preferential heating of electrons through EBWs generates (collisionless) energetic electrons (b) which creates an asymmetry amongst the collisional thermal electrons. Collisions lead to the restoration of a Maxwellian amongst the thermal electrons (c) such that the thermal electrons carry no current, and the current is carried by the energetic electrons. In reality, collisions are not sufficiently strong to fully restore the Maxwellian, and the EBW heating will flatten the distribution in $p_\perp$, as shown in figure \ref{fig:startup:collisions}(b), rather than create a well-defined group of energetic electrons.}
	\label{fig:startup:collisions:theory}
	\end{figure}

The current generated by the two different values of $N_\parallel$ illustrates the Fisch-Boozer mechanism: the preferential heating of electrons, with $N_\parallel = 0.5$ heats electrons with $p_\parallel > 0$, generating a positive current, while $N_\parallel = 0$ fails to gain a directionality with respect to the magnetic field, and therefore does not generate a plasma current. A schematic of the Fisch-Boozer CD mechanism is shown in figure \ref{fig:startup:collisions:theory}.

This simple test confirms several EBW CD theories. Firstly, EBWs with $N_\parallel = 0$, as is the case for EBWs propagating close to the midplane, do not contribute significantly to the CD as it fails to gain a directionality with respect to the magnetic field \cite{Forest_2000, Urban_2011}. Secondly, by creating an EBW with optimal $N_\parallel$, gaining a directionality with respect to the magnetic field, a current can be generated \cite{Shevchenko_2002}, but due to the formation of energetic electrons, this current is small, excluding the Fisch-Boozer mechanism as a major contributor to the CD \cite{Maekawa_2005}.

\newpage
\subsection{Current drive by the preferential confinement of electrons}
Wong \cite{Wong_1980} observed that during start-up when the magnetic field lines have an open configuration, a preferential confinement of electrons can be created by adding a small vertical magnetic field. This preferential confinement has been used to describe the initiation of CFS, using single particle orbits \cite{Ejiri_2007, Yoshinaga_2006, Maekawa_2012}, and it was shown in Section \ref{sec:loss} that until CFS form, the confinement of forward electrons is much better than electrons moving counter to the magnetic field.

\begin{figure}[!hbt]
	\centering
		\subfloat[]{%
			\includegraphics[width=0.33\textwidth]{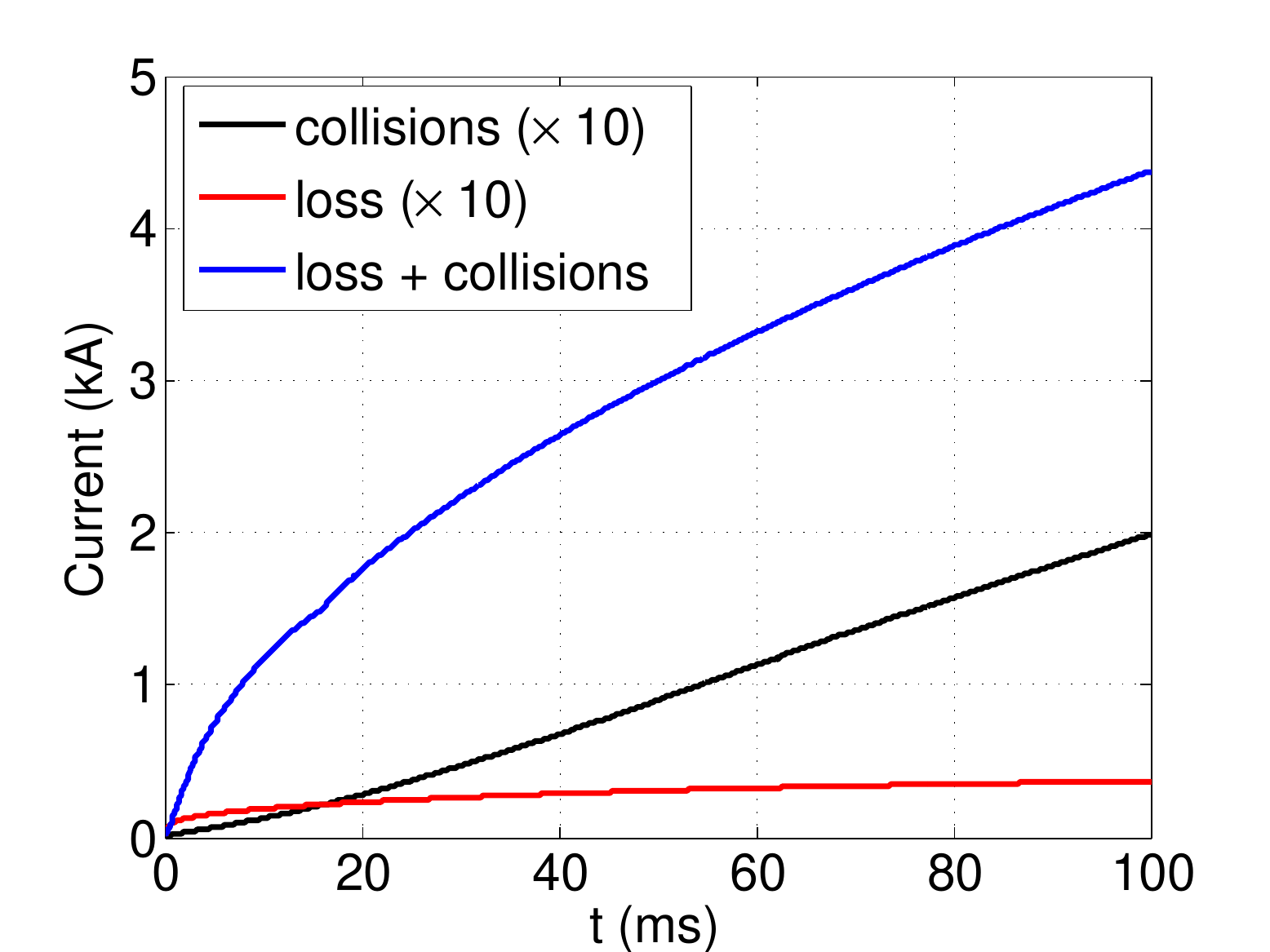}}
		\hfill
		\subfloat[]{%
			\includegraphics[width=0.33\textwidth]{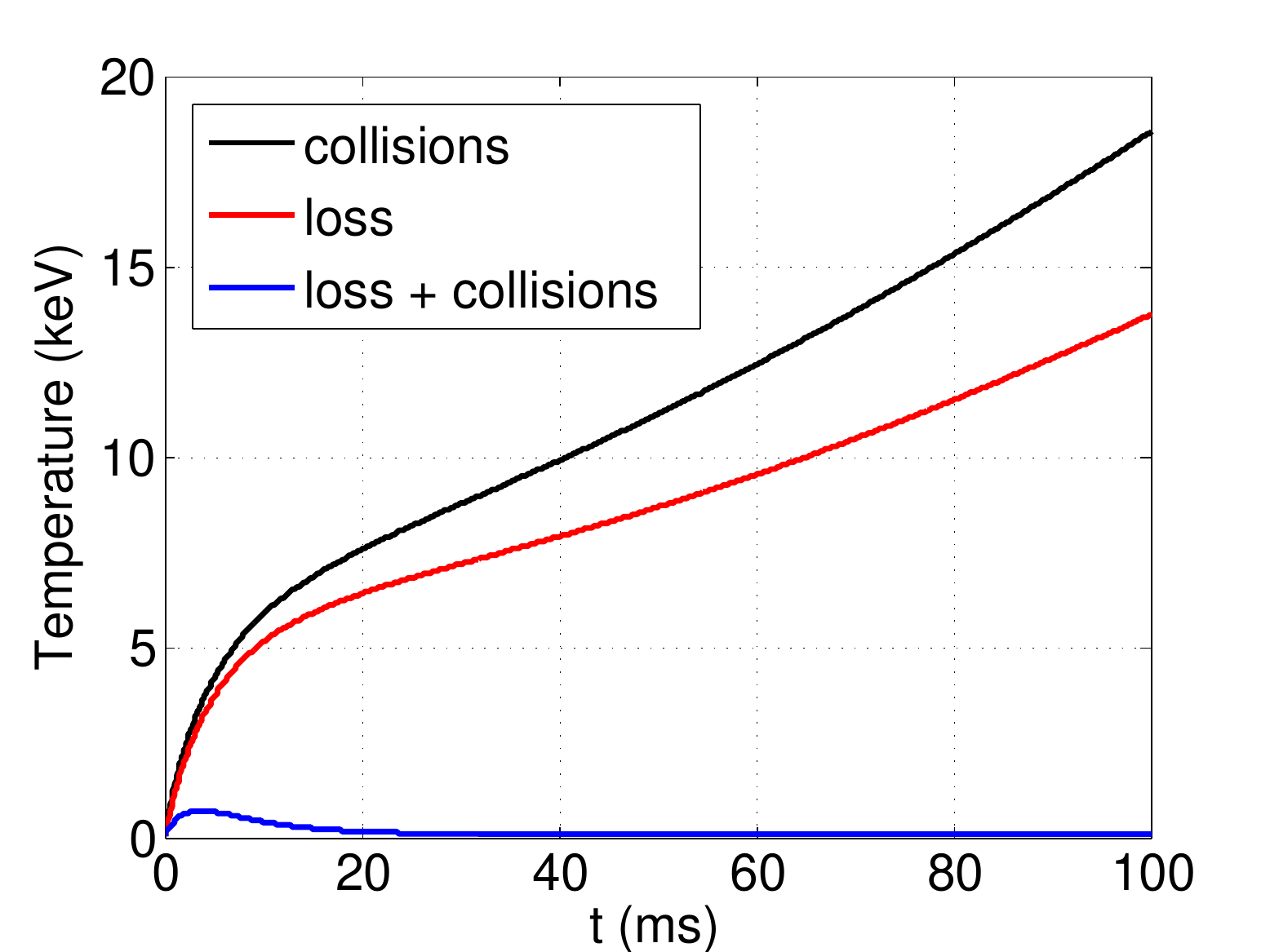}}
		\hfill
		\subfloat[]{%
			\includegraphics[width=0.33\textwidth]{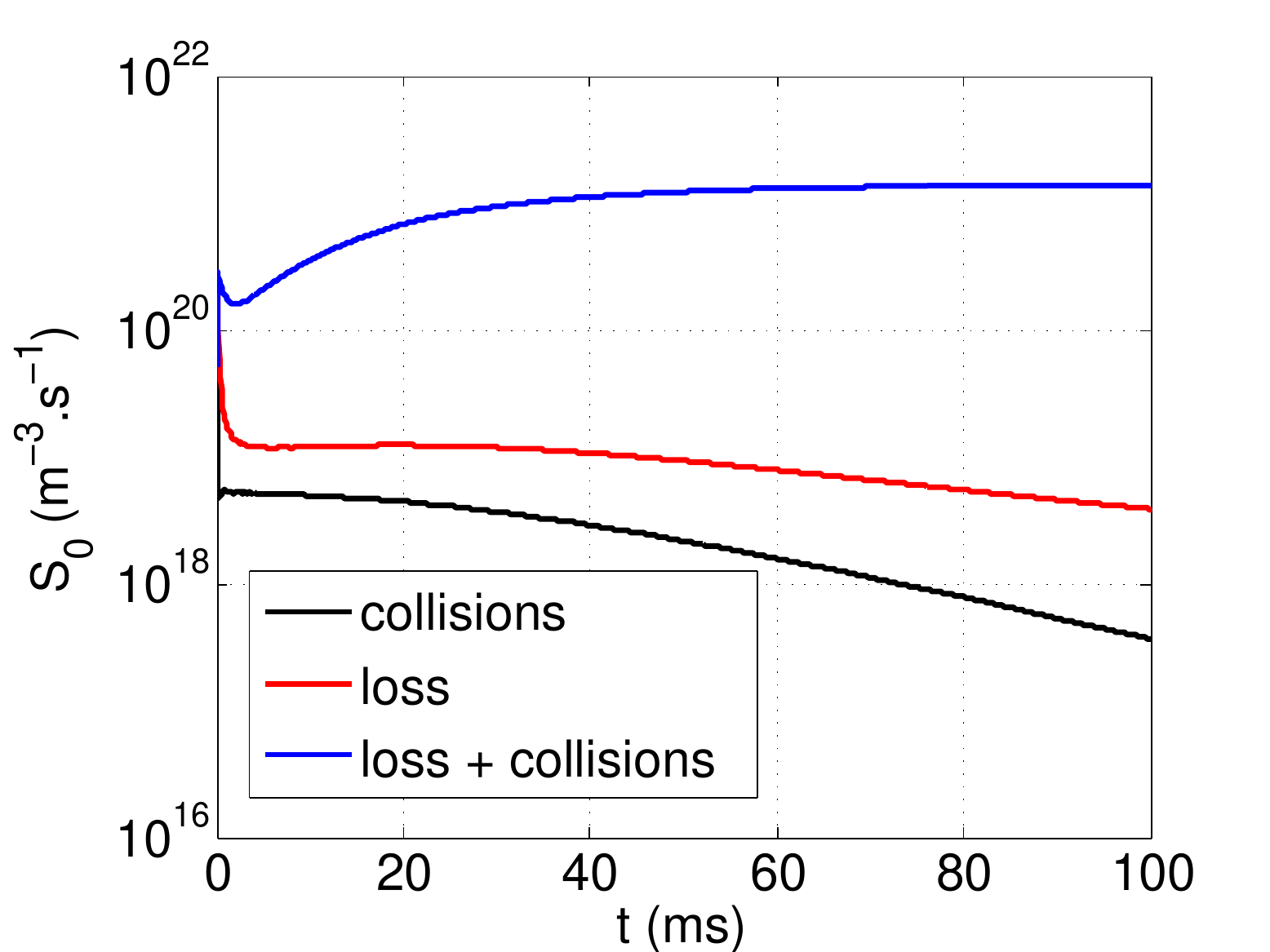}}
	\caption[]{The time evolution of (a) the plasma current shows that a significantly larger current is generated when both losses and collisions are present, indicative of the current drive mechanism through the preferential confinement of electrons. The time evolution of (b) the temperature also shows that by including the loss term, the electron temperature is significantly lower, even though all cases have the same density. Lost electrons are replaced with cold electrons through the source, and the time evolution of (c) $S_0$ shows that the number of cold electrons ($T_e = 2 \, \textrm{eV}$) added to the system is about $3$ orders of magnitude more.}
	\label{fig:startup:collisions_loss}
	\end{figure}

In order to test the preferential confinement of electrons as a possible CD mechanism, consider a start-up simulation,
	\begin{equation*}
	\frac{\partial f}{\partial t} = \textrm{source} + \textrm{RF heating} + \textrm{loss} + \textrm{loop voltage} +  \textrm{collisions} 
	\end{equation*}
where the vacuum vertical field $B_V = 6 \, \textrm{mT}$ is fixed. In order to see the effect of the loss term on the generated current, we perform simulations with and without collisions, and compare this to the case with no loss term. From the previous section, we know that the preferential heating of electrons is necessary to generate a current, so we set $N_\parallel = 0.5$ and $\Delta R = 0.05 \, \textrm{m}$ in equation (\ref{eq:EBW:dR}). Results are shown in figure \ref{fig:startup:collisions_loss}.

\begin{figure}[!hbt]
	\centering
	\includegraphics[width=1\textwidth]{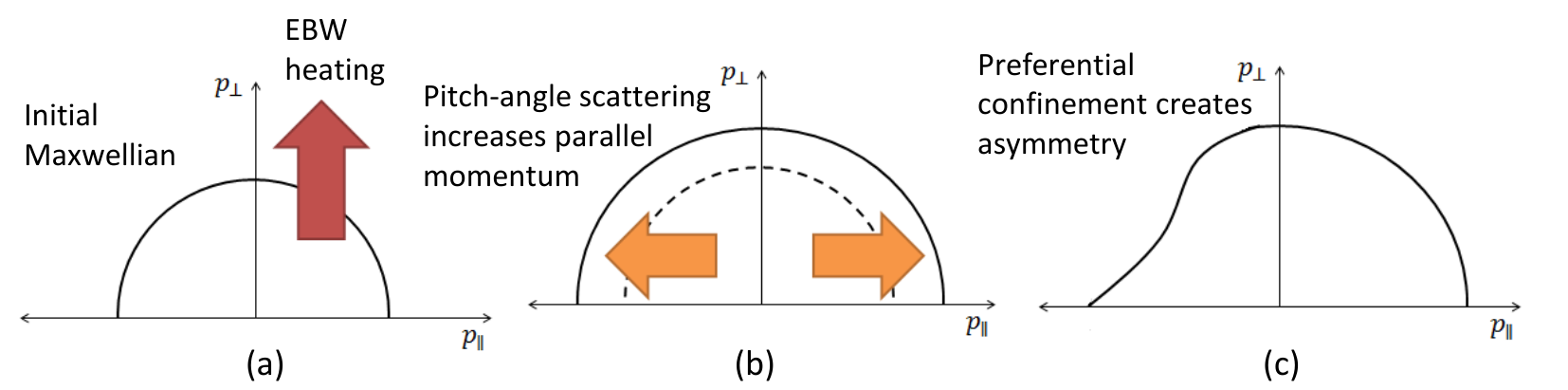}
	\caption[]{Schematic of the current drive mechanism through the preferential confinement of electrons. (a) EBW heating increases the temperature of the distribution and (b) the parallel momentum of electrons through pitch-angle scattering. (c) The preferential confinement of electrons with $p_\parallel > 0$, coupled with the greater loss of electrons with $p_\parallel < 0$, leads to an increase in the generated current.}
	\label{fig:startup:collisions_loss:theory}
	\end{figure}

The comparison of the simulated plasma current in three different scenarios is shown in figure \ref{fig:startup:collisions_loss}(a). In the absence of collisions, the plasma current generated by the loss term is even smaller than the current generated by the Fisch-Boozer mechanism. As the EBW heating increases only the perpendicular momentum of electrons, which does not generate a current in itself, the losses of electrons are small. Including collisions allows the parallel momentum of electrons to be increased through pitch-angle scattering, leading to greater losses and a generated current more than $10$ times greater than before. The preferential confinement of electrons is therefore responsible for the greater part of the generated current, with collisions only ``feeding''the loss term by increasing the parallel momentum of electrons through pitch-angle scattering. A schematic of the mechanism is shown in figure \ref{fig:startup:collisions_loss:theory}.

By relating the average energy in each distribution to a pseudo-temperature, we can compare the ``temperatures'' of the three cases, shown in figure \ref{fig:startup:collisions_loss}(b). We note that, without any losses, a very high temperature is reached. All three simulations have the same density and EBW power absorbed, such that, in the case with no losses, all the absorbed power is used to increase the temperature. When adding losses, but excluding collisions, the temperature is slightly lower, as very few electrons are lost, due to EBW heating increasing only the perpendicular momentum of electrons. With both collisions and orbital losses, the temperature is greatly reduced, to the order of hundreds of $\textrm{eV}$, which is comparable to temperatures inferred from experiments. In this case, collisions increase the parallel momentum of electrons through pitch-angle scattering, ``feeding'' the loss term. Fast electrons are therefore lost, and, in order to ensure the density remains the same, are replaced by cold electrons, leading to a reduction in the electron temperature.

Figure \ref{fig:startup:collisions_loss}(c) show the electron injection rate for the three different cases considered. As all three cases have the same time evolution of density, the value of $S_0$ needs to be adjusted depending on the number of electrons lost. Adding the loss term leads to an increase in the value of $S_0$, while having both collisions and orbital losses lead to an even greater increase, as more electrons are lost.

The current drive mechanism is therefore based on the preferential confinement of electrons, with collisions increasing the loss rate through pitch-angle scattering.

\newpage
\subsection{Vertical shift}
Experiments conducted on MAST indicated that shifting the plasma up or down helps the formation of CFS \cite{Shevchenko_2010, Shevchenko_2015}. Shifting the plasma a distance $Z_0$ influences start-up in two ways:

	\begin{enumerate}

	\item It influences the confinement of electrons, enhancing the asymmetry, as was shown in Section \ref{sec:loss}, in such a way that a larger current can be generated.

	\item It generates a favourable $B_V$ in the MC zone in order for an optimal $N_\parallel$ to develop. In MAST, $B_V < 0$ which leads to $N_\parallel < 0$ above the midplane and $N_\parallel > 0$ below the midplane. In order to have $N_\parallel > 0$ in the MC zone, the plasma is shifted upwards. Once CFS start to form, $B_Z$ changes sign in the MC region, and the plasma has to be shifted back downwards to ensure $N_\parallel > 0$ in the MC zone.

	\end{enumerate}

\begin{figure}[!hbt]
	\centering
		\hspace*{\fill}
			\includegraphics[width=0.33\textwidth]{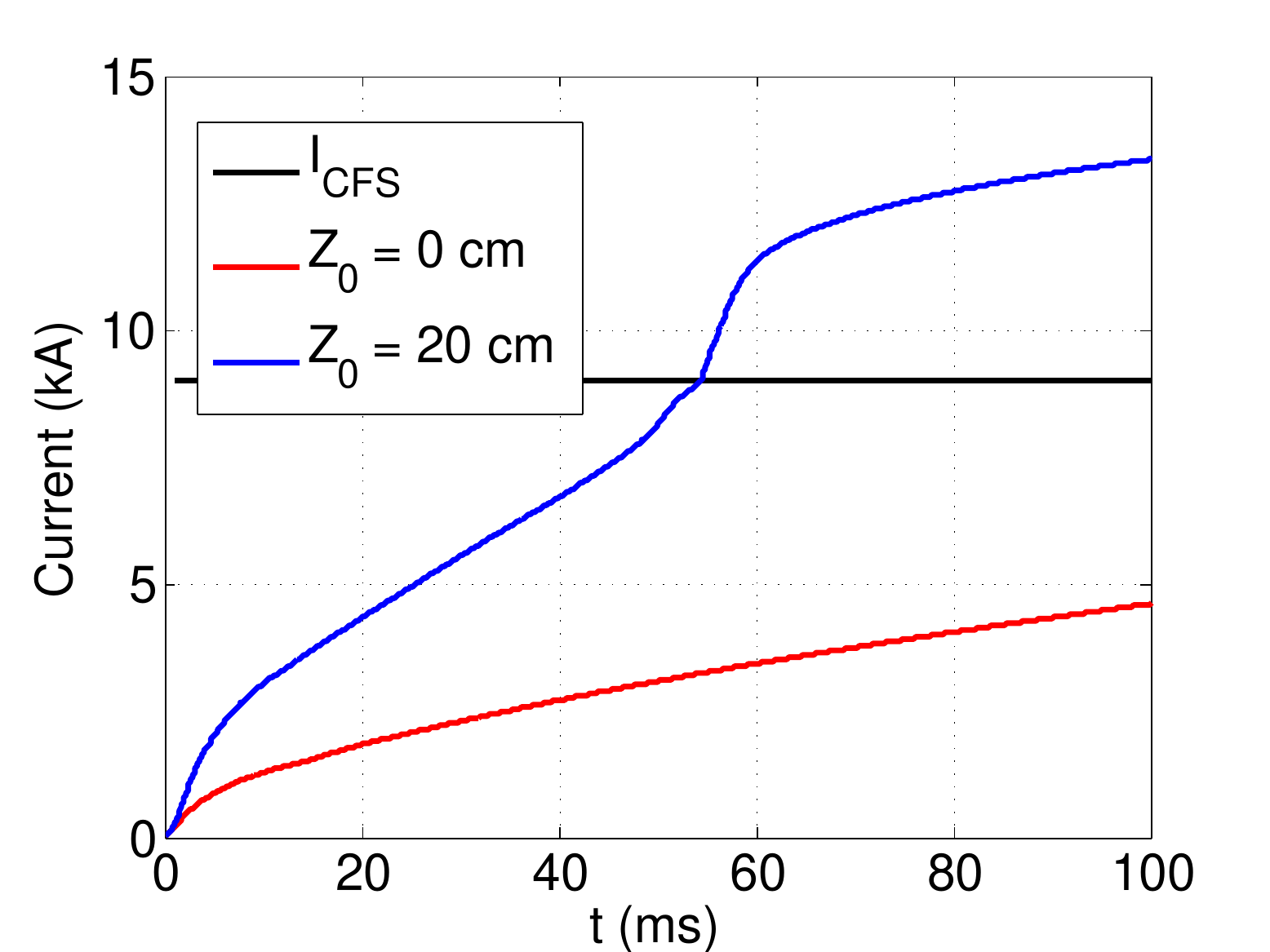}
		\hspace*{\fill}
	\caption[]{Shifting the plasma upwards, to create a favourable value of $N_\parallel$, generates a greater plasma current compared to not shifting the plasma, helping the formation of CFS.}
	\label{fig:startup:Z0}
	\end{figure}

Through experiments and ray-tracing we know that the value of $N_\parallel$ changes as the EBW propagates towards the ECR layer. In order to account for this, we let $\Delta N_\parallel = 1$ in (\ref{eq:EBW:dN}), and compare two cases: $Z_0 = 20 \, \textrm{cm}$ with $N_\parallel = 0.5$, and $Z_0 = 0 \, \textrm{cm}$ with $N_\parallel = 0$. We let the vacuum magnetic field be constant, leading to a constant value of $I_\textrm{\small{CFS}}$. The comparison of the generated plasma current, for an input power $P_0 = 50 \, \textrm{kW}$, is shown in figure \ref{fig:startup:Z0}, showing a large increase in current, and the formation of CFS, when shifting the plasma upwards, as observed experimentally.

\subsection{Vacuum field ramp-up}
Experiments showed that the most efficient method of generating larger plasma currents is by increasing the vacuum field strength \cite{Shevchenko_2010, Shevchenko_2015}. In Section \ref{sec:loss} we showed that increasing the vacuum field strength leads to an increase in the value of $I_\textrm{\small{CFS}}$, the value of the plasma current where the first CFS start to form. An increase in $I_\textrm{\small{CFS}}$ can lead to a subsequent increase in $I_P$ while keeping the asymmetry of the loss term intact, generating larger plasma currents.

\begin{figure}[!hbt]
	\centering
		\hspace*{\fill}
		\subfloat[]{%
			\includegraphics[width=0.33\textwidth]{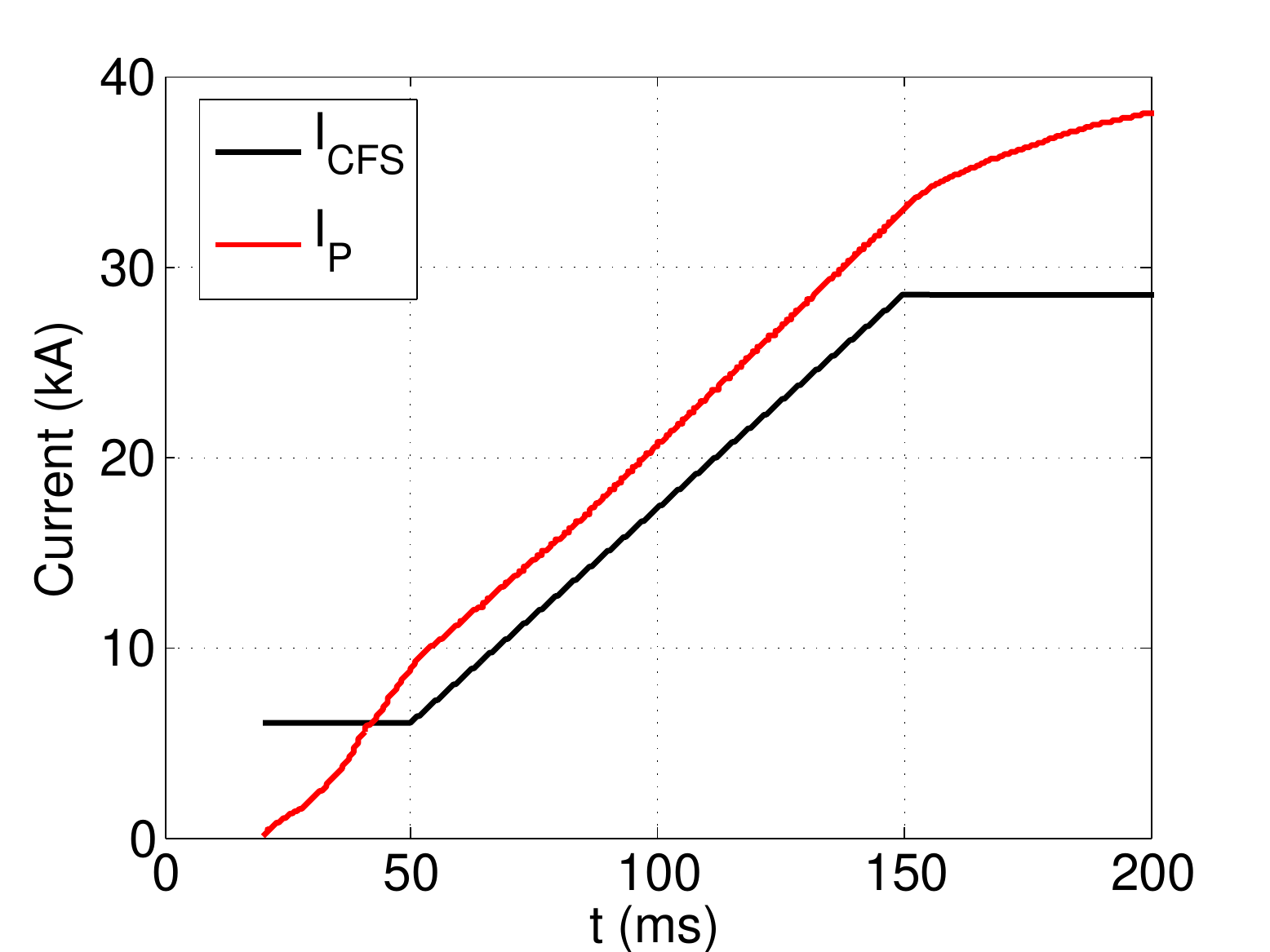}}
		\hfill
		\subfloat[]{%
			\includegraphics[width=0.33\textwidth]{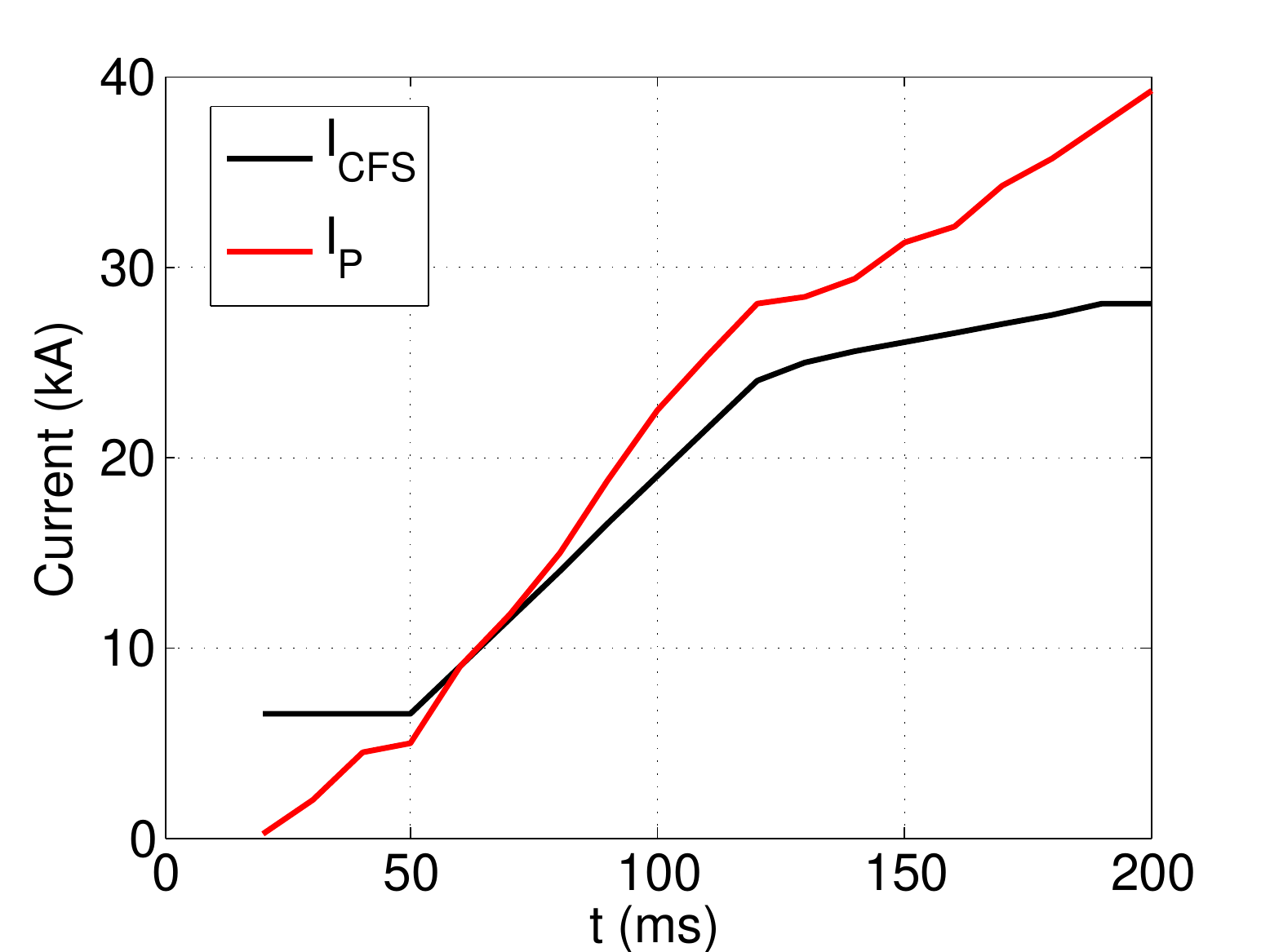}}
		\hspace*{\fill}
	\caption[]{The best way of generating a larger plasma current is by increasing the vacuum field strength, or the value of $I_\textrm{\small{CFS}}$, as illustrated here for (a) a simulated plasma current, compared to $I_\textrm{\small{CFS}}$, and (b) experimentally measured plasma current in MAST shot $\# 28941$. By increasing $I_\textrm{\small{CFS}}$, the plasma current $I_P$ can be increased while sustaining the asymmetry in the loss term, leading to larger plasma currents.}
	\label{fig:startup:ramp}
	\end{figure}

In order to test this, we replicate the magnetic field strength and vertical shift of an experiment conducted on MAST. The vertical shift ensures that $N_\parallel > 0$, while also affecting the confinement of electrons, to help the formation of CFS. We therefore let $N_\parallel = 0.5$ and $\Delta N_\parallel = 1$ in equation (\ref{eq:EBW:dN}). Results for a $50 \, \textrm{kW}$ input is shown in figure \ref{fig:startup:ramp}, and shows a rise in $I_P$ as $I_\textrm{\small{CFS}}$ is increased, generating larger plasma currents, due to the asymmetry of the loss term being maintained through the increase in $I_\textrm{\small{CFS}}$.

The rise in plasma current compares well to experiment, as the current generated follows the increase in $I_\textrm{\small{CFS}}$. As the vacuum field strength is ramped-up, and the value of $I_\textrm{\small{CFS}}$ is increased, the plasma current $I_P$ can increase accordingly, as the asymmetry in the loss term is sustained.

\begin{figure}[!hbt]
	\centering
		\hspace*{\fill}
		\subfloat[]{%
			\includegraphics[width=0.33\textwidth]{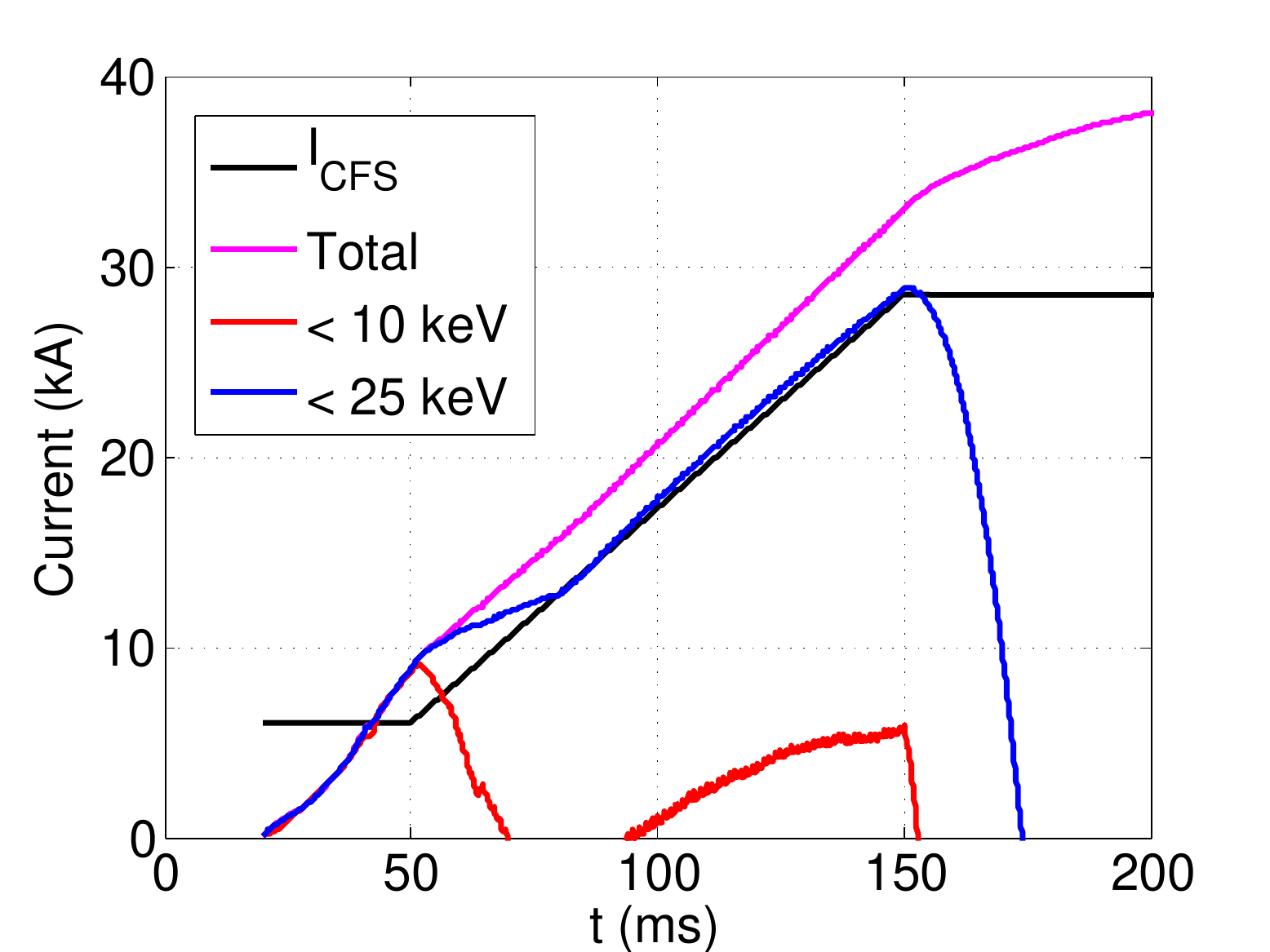}}
		\hfill
		\subfloat[]{%
			\includegraphics[width=0.33\textwidth]{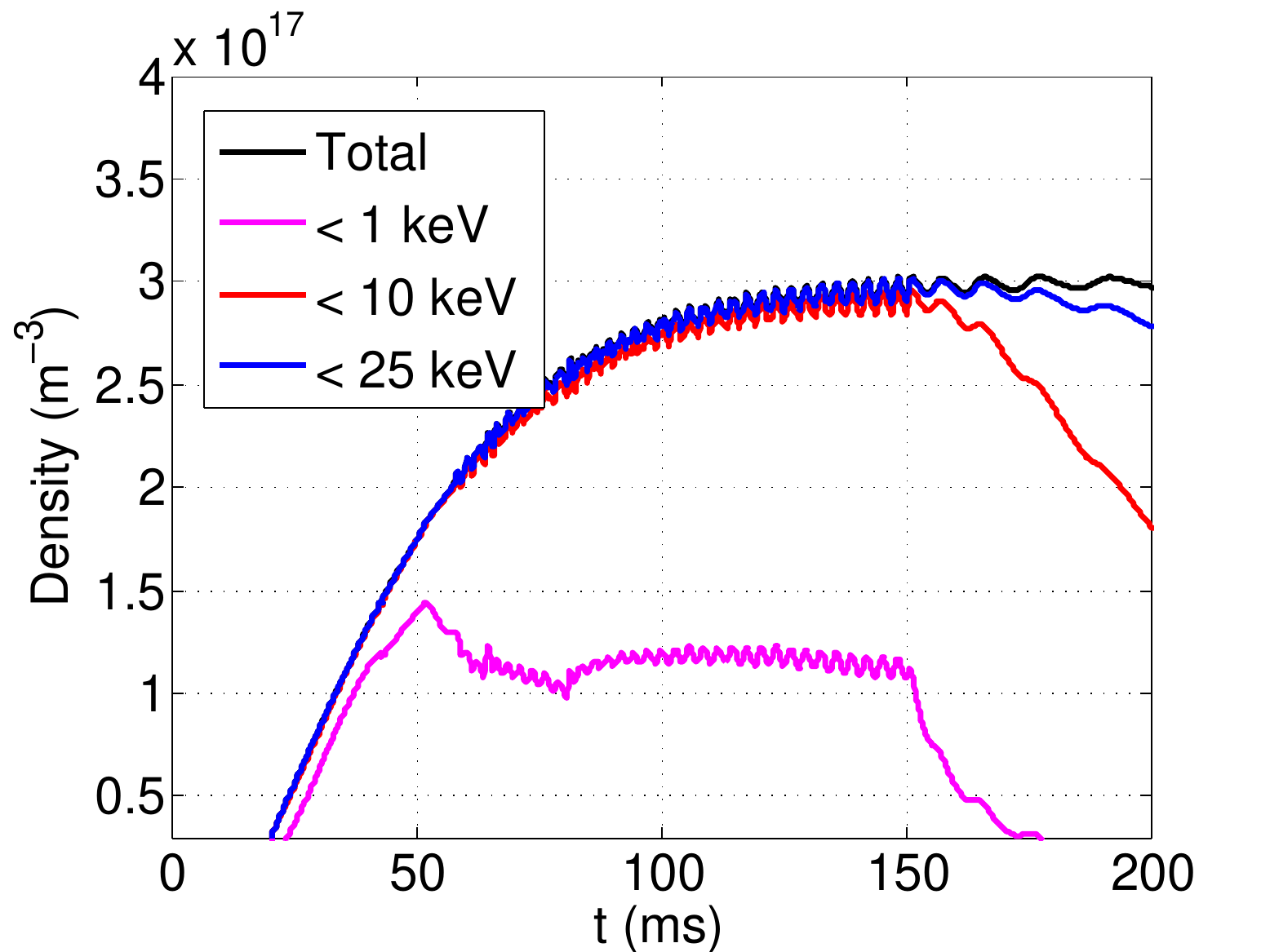}}
		\hspace*{\fill}
	\caption[]{Experiments concluded that the majority of the plasma current is carried by energetic electrons \cite{Shevchenko_2010, Shevchenko_2015, Maekawa_2005, Yoshinaga_2006}, which is confirmed by simulation. The time evolution of (a) the plasma current, shows that the majority of current is carried by electrons between $10-25 \, \textrm{keV}$ during the vacuum field ramp-up, with this energy increasing after CFS start to form. Electrons with $v_\parallel > 0$ are accelerated by the EBW, such that the thermal electrons carry a negative current. Although the current is carried by these energetic electrons, the majority of electrons have energies less than $10 \, \textrm{keV}$, as is shown in (b) the time evolution of the electron density.}
	\label{fig:startup:energy}
	\end{figure}

Experiments further concluded that the majority of the plasma current is carried by energetic electrons \cite{Shevchenko_2010, Shevchenko_2015, Maekawa_2005, Yoshinaga_2006}, which raises the issue of whether a collision-driven CD mechanism is valid. Figure \ref{fig:startup:energy} shows that the majority of the plasma current is carried by electrons with energies $10-25 \, \textrm{keV}$ during the vacuum field ramp-up phase, while the majority of electrons have energies less than $10 \, \textrm{keV}$. This compares well to experimental conclusions, which suggests that electrons with energies $\sim 25 \, \textrm{keV}$ are responsible for the majority of the current, but that the majority of electrons have energies much lower.

As electrons with $v_\parallel > 0$ are accelerated by the EBW, the thermal electrons carry a negative current, as there are more electrons with $v_\parallel < 0$ than $v_\parallel > 0$. The energetic electrons with energies $10-25 \, \textrm{keV}$ compensate for this negative current with a larger positive current.

After the vacuum field ramp-up phase, the energies of the electrons responsible for carrying the majority of the plasma current increases, as the confinement of electrons improves and collisions become responsible for the current drive, in a situation similar to that depicted in figure \ref{fig:startup:collisions:theory}.

\subsection{The relationship between power, density and current}
Experiments on MAST found that there exists a linear relationship between the generated plasma current and injected RF power, with an efficiency of about $1 \, \textrm{A}/\textrm{W}$. In order to gain an understanding of this relationship, it is necessary to first look at the dependence of the generated current on density and power, as the injected power will have an effect on the electron density, which will then impact the plasma current.

\begin{figure}[!hbt]
	\centering
		\hspace*{\fill}
		\subfloat[]{%
			\includegraphics[width=0.33\textwidth]{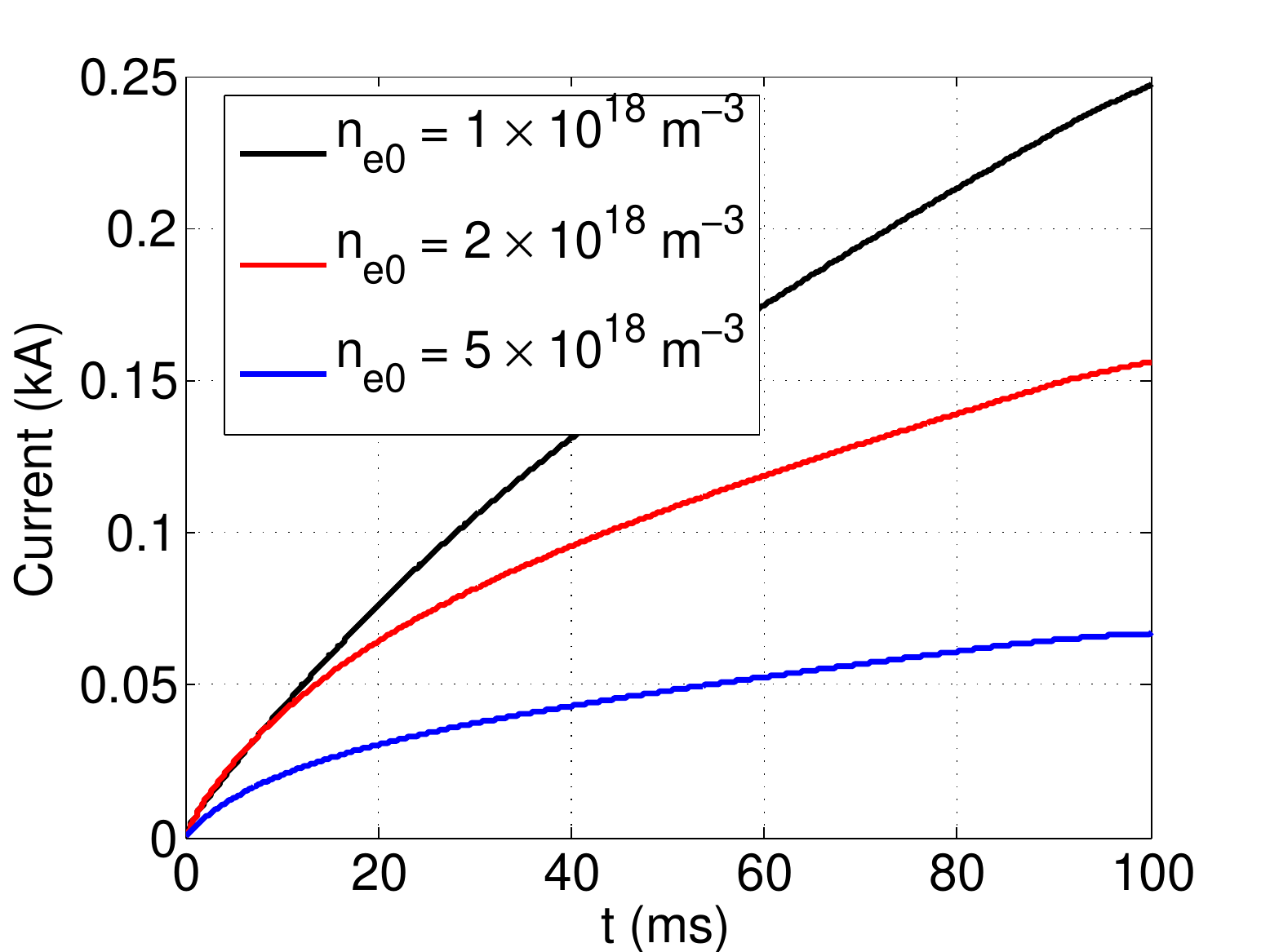}}
		\hfill
		\subfloat[]{%
			\includegraphics[width=0.33\textwidth]{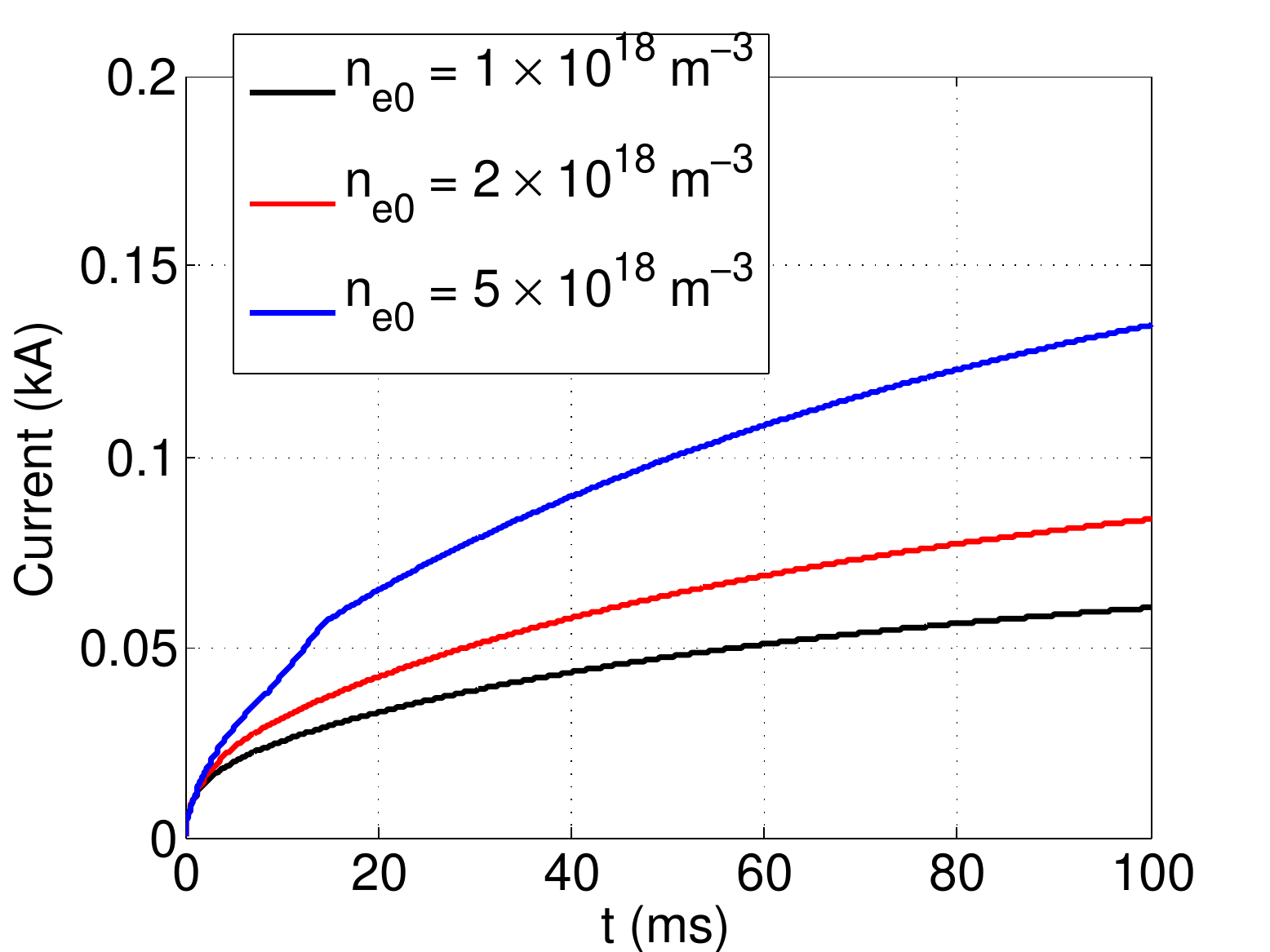}}
		\hfill
		\subfloat[]{%
			\includegraphics[width=0.33\textwidth]{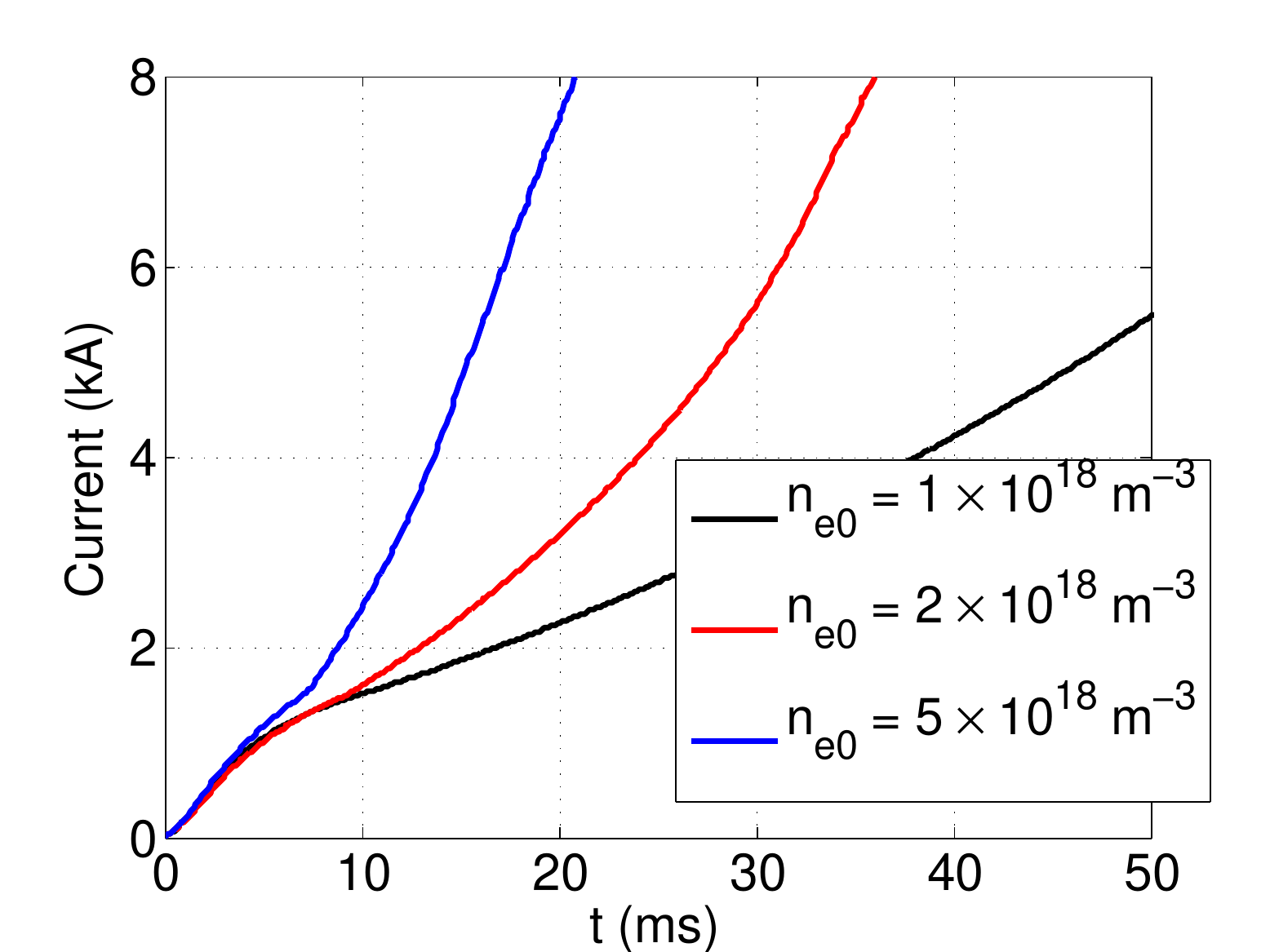}}
		\hspace*{\fill}
	\caption[]{Comparison of the generated plasma current under (a) collisions (no losses), (b) losses (no collisions), and (c) both losses and collisions, for increasing electron density and constant power. The collisional current decreases when increasing the density, while the current generation by the preferential confinement of electrons increases for increasing density.}
	\label{fig:startup:density}
	\end{figure}

The current generated by the Fisch-Boozer mechanism decreases for increasing electron density, as shown in figure \ref{fig:startup:density}(a). For the same power absorbed, higher density distributions will have lower temperatures and increased collisionality. This will force the distribution to be closer to Maxwellian, thereby reducing the anisotropy in the plasma resistivity.

The current generated by the preferential confinement of electrons, however, increases as the density is increased, as shown in figure \ref{fig:startup:density}(b). As electrons with $v_\parallel < 0$ are lost faster than electrons with $v_\parallel > 0$, increasing the density will result in greater losses, and therefore a greater plasma current.

Figure \ref{fig:startup:density}(c) shows the evolution of plasma current when including both collisions and orbital losses for different electron densities. The increase in density results in greater losses, which leads to a greater plasma current, similarly to the case without collisions, and the generated current is proportional to the electron density.

\begin{figure}[!hbt]
	\centering
		\hspace*{\fill}
		\subfloat[]{%
			\includegraphics[width=0.33\textwidth]{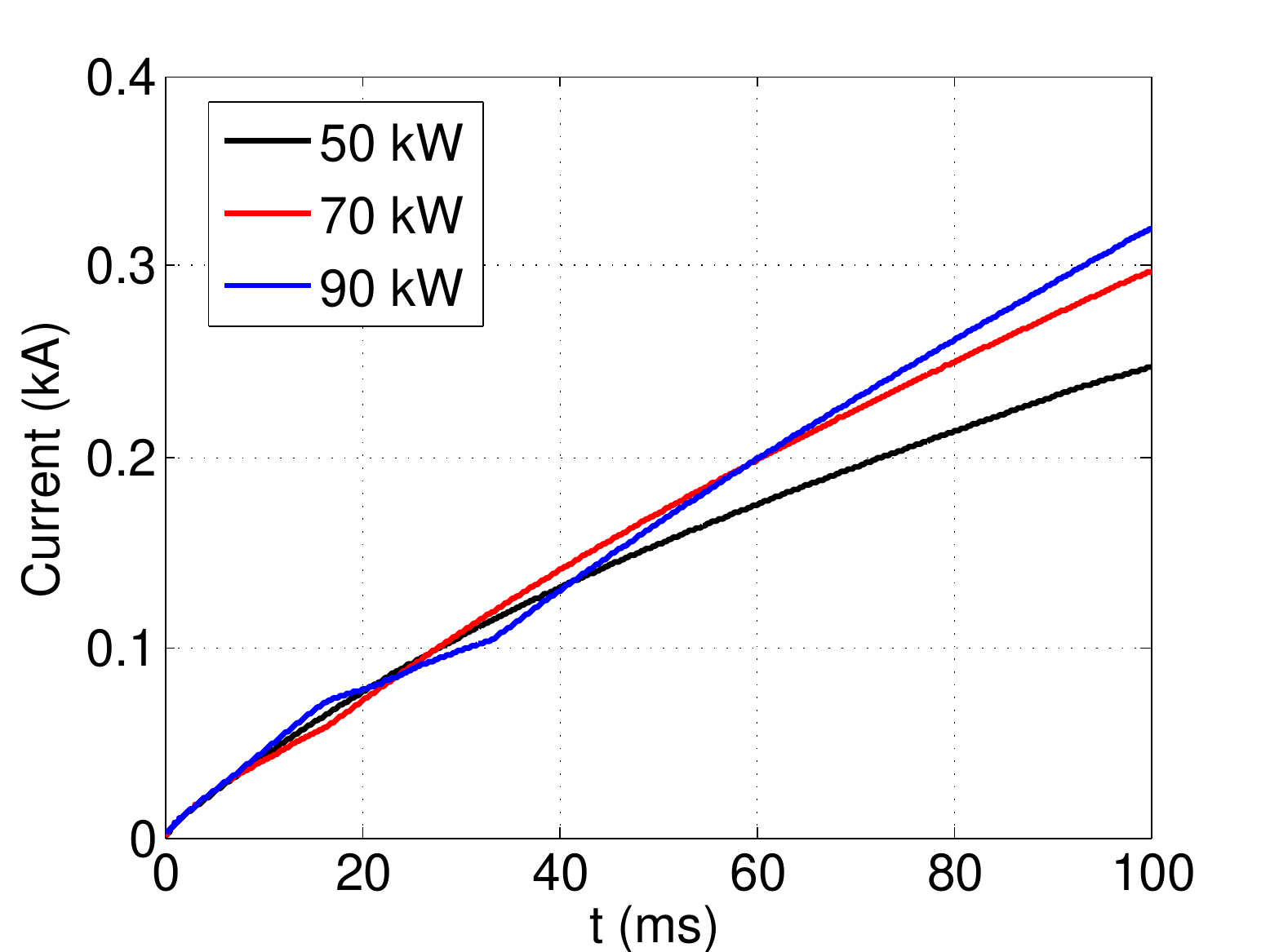}}
		\hfill
		\subfloat[]{%
			\includegraphics[width=0.33\textwidth]{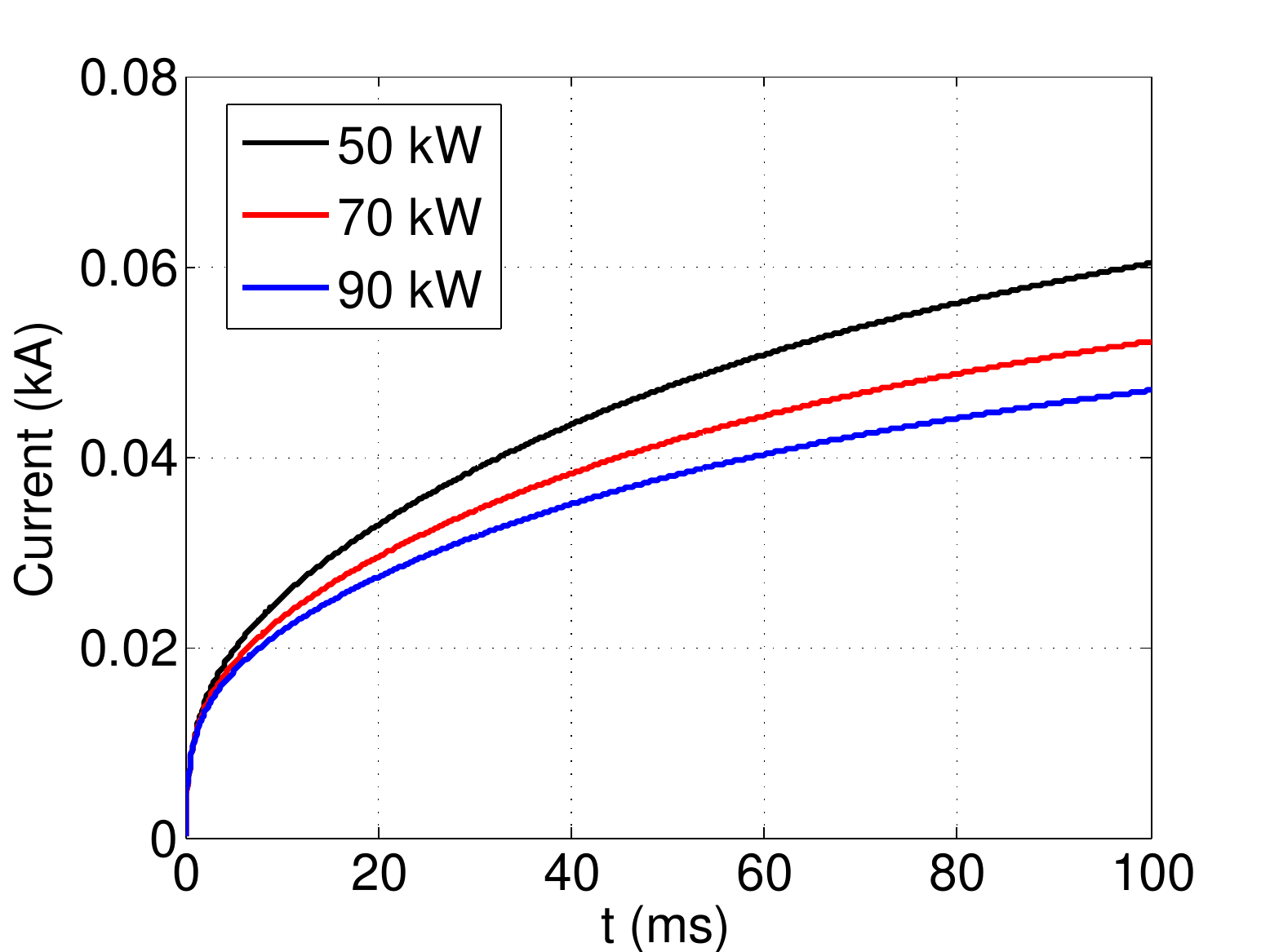}}
		\hfill
		\subfloat[]{%
			\includegraphics[width=0.33\textwidth]{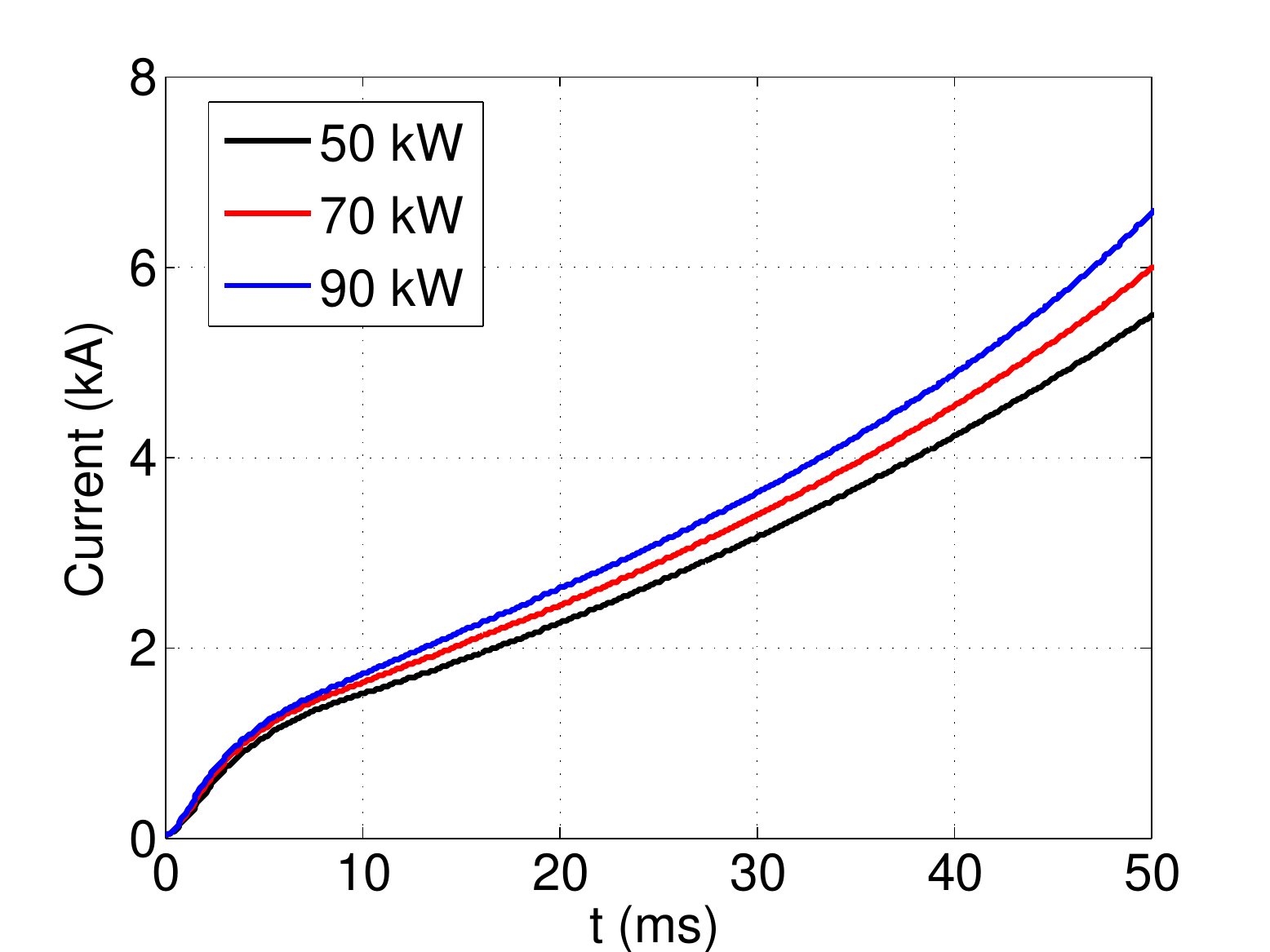}}
		\hspace*{\fill}
	\caption[]{Comparison of the generated plasma current under (a) collisions (no losses), (b) losses (no collisions), and (c) both losses and collisions, for increasing RF power and constant density. The current generated by both collisions by the preferential confinement of electrons, with collisions, is increases for increasing power, while it decreases with increasing power in the absence of collisions.}
	\label{fig:startup:power}
	\end{figure}

The dependence of plasma current on injected power is shown in figure \ref{fig:startup:power}. In this case, the electron density is kept constant, and the power absorbed is varied. The current generated by the Fisch-Boozer mechanism is shown to increase when increasing the absorbed power, as higher power will lead to higher temperatures and decreased collisionality, leading to a greater anisotropy in the plasma resistivity.

In the absence of collisions, the current generated by the preferential confinement of electrons decreases for increasing power, but increases when collisions are included. As EBW power mainly increases the perpendicular momentum of electrons, increasing the power will create more electrons with large $p_\perp$, which reduces the probability of an electron being lost. In the absence of collisions, electrons can not be pitch-angle scattered into regions of larger $p_\parallel$ and greater losses, such that the overall losses decrease, and the generated plasma current is smaller.

When including both losses and collisions, increasing the power leads to an increase in the temperature, including the parallel temperature. Therefore, electrons have, on average, larger values of $p_\parallel$, and, as orbital losses increase for increasing $p_\parallel$, increasing the power leads to greater losses and larger plasma currents.

In order to fully understand how the injected power and electron density influences the generated plasma current, a better understanding of the relationship between power and density is required. While experiments on MAST showed a linear relationship between the plasma current and injected power \cite{Shevchenko_2015}, Yoshinaga \cite{Yoshinaga_2006} observed that, for a fixed vacuum magnetic field strength, increasing the injected power does not increase the total generated current, but only decreases the amount of time it takes for CFS to form. Gaining a better understanding of these observations is not possible without knowledge of the density evolution as a function of input power, and such a study is beyond the scope of this work.

\section{Discussion}
EBW-assisted plasma current start-up has previously been demonstrated successfully in a number of experiments. An important aspect of start-up is the change in the magnetic field topology, from an open field line configuration to the formation of CFS, which is governed by the initiation of a plasma current. The generated current is mainly driven by the absorption of EBW power, and carried by energetic electrons, while larger currents are generated through an increase of the vacuum magnetic field strength. 

Previous studies into the current drive mechanism have focused mainly on pressure-driven currents and the study of single particle orbits for the initiation of CFS through a preferential confinement of electrons based on the open magnetic field line structure. This paper describes the development of a kinetic model to simulate the dynamic start-up phase, and shows that collisions are responsible for only a small part of current drive, while the preferential confinement of electrons, controlled by the vacuum magnetic field, are responsible for the majority of the driven current.

The Fisch-Boozer mechanism, based on the preferential heating of electrons to create an anisotropic plasma resistivity, generates plasma currents much smaller than those observed in experiments. The open magnetic field line configuration leads to the preferential confinement of electrons, due to the vertical magnetic field cancelling out the perpendicular $\nabla B$ and curvature drifts for a selection of electrons. Collisions then act to ``feed'' the loss term: RF heating mainly increases the perpendicular momenta of electrons, and through pitch-angle scattering the parallel momenta of electrons are increased, increasing the rate at which electrons are lost, and ultimately leading to larger plasma currents being generated.

The location of RF heating, in this case through EBWs, depends on the local magnetic field in the MC zone. In MAST, the MC zone is located just above the midplane, and, to ensure $N_\parallel > 0$, the magnetic field midplane must be shifted upwards. Once CFS start to form, however, the poloidal magnetic field reverses direction in the MC zone, and the magnetic field midplane must be shifted back downwards to ensure $N_\parallel > 0$ remains. This vertical shift also influences the confinement of electrons, and helps the formation of CFS.

The preferential confinement of electrons is an effect that disappears once CFS are completely formed and all electrons are confined. It can, however, be controlled through the use of the vacuum magnetic field in order to generate larger plasma currents. If we consider the plasma current $I_P = I_\textrm{\small{CFS}}$ to be the point where all forward electrons are confined and the first CFS start to form, then by increasing the value of $I_\textrm{\small{CFS}}$, larger values of $I_P$ can be achieved while keeping the asymmetry of the loss term intact. In this way a $B_V$ ramp-up can be used to achieve greater plasma currents.

Simulations show that the majority of the plasma current is carried by electrons with energies $10-25 \, \textrm{keV}$, which compares well to experimental conclusions. After CFS forms, however, the energies of current carrying electrons start to increase, as the distribution function becomes flat and electrons are accelerated to higher and higher energies, as depicted in figure \ref{fig:startup:collisions:theory}.

Understanding the dependence of the plasma current on the injected power requires knowledge of the time evolution of the electron density, and its dependence on the injected power, which is beyond the scope of this work. It is shown, however, that the current generated by the Fisch-Boozer mechanism decreases for increasing electron density, and increases when increasing the absorbed power, while the current generated by the preferential confinement of electrons increases when increasing both the electron density and absorbed power.

Losses and collisions have the net effect of producing a current which increases with RF power, as observed experimentally, but the model does not give the linear scaling of current with power as seen on MAST. This relationship may be reproduced through further investigation and refinements of the model.

\section*{Acknowledgements}
This work is funded by the RCUK Energy Programme under grant EP/P012450/1, and the University of York, through the Department of Physics and the WW Smith Fund.


\section*{References}

\begin{thebibliography}{}

\bibitem{Forest_1994}
C.B. Forest \textit{et al.}, \textit{Phys. Plasmas} \textbf{1}, 1568 (1994).

\bibitem{Ejiri_2006}
A. Ejiri \textit{et al.}, \textit{Nucl. Fusion} \textbf{46}, 709 (2006).

\bibitem{Uchida_2004}
M. Uchida \textit{et al.}, \textit{J. Plasma Fusion Res.} \textbf{80}, 83 (2004).

\bibitem{Shevchenko_2007}
V.F. Shevchenko \textit{et al.}, \textit{Fusion Sci. Technol.} \textbf{52}, 202 (2007).

\bibitem{Shevchenko_2010}
V.F. Shevchenko \textit{et al.}, \textit{Nucl. Fusion} \textbf{50}, 022004 (2010).

\bibitem{Maekawa_2005}
T. Maekawa \textit{et al.}, \textit{Nucl. Fusion} \textbf{45}, 1439 (2005).

\bibitem{Shevchenko_2015}
V.F. Shevchenko \textit{et al.}, \textit{EPJ Web of Conf.} \textbf{87}, 02007 (2015).

\bibitem{Fisch_1980}
N.J. Fisch and A.H. Boozer, \textit{Phys. Rev. Lett.} \textbf{45}, 720 (1980).

\bibitem{Forest_2000}
C.B. Forest, P.K. Chattopadhyay, R.W. Harvey and A.P. Smirnov, \textit{Phys. Plasmas} \textbf{7}, 1352 (2000).

\bibitem{Shevchenko_2002}
V.F. Shevchenko, Y. Baranov, M. O'Brien and A. Saveliev, \textit{Phys. Rev. Lett.} \textbf{89}, 265005 (2002).

\bibitem{Petrillo_1987}
V. Petrillo, G. Lampis and C. Maroli, \textit{Plasma Phys. Control. Fusion} \textbf{29}, 877 (1987).

\bibitem{Yoshinaga_2007}
T. Yoshinaga, M. Uchida, H. Tanaka and T. Maekawa, \textit{Nucl. Fusion} \textbf{47}, 210 (2007).

\bibitem{Maekawa_2012}
T. Maekawa, T. Yoshinaga, M. Uchida, F. Watanabe and H. Tanaka, \textit{Nucl. Fusion} \textbf{52}, 083008 (2012).

\bibitem{Ejiri_2007}
A. Ejiri and Y. Takase, \textit{Nucl. Fusion} \textbf{47}, 403 (2007).

\bibitem{Yoshinaga_2006}
T. Yoshinaga, M. Uchida, H. Tanaka and T. Maekawa, \textit{Phys. Rev. Lett.} \textbf{96}, 125005 (2006).

\bibitem{Kim_2012}
H.T. Kim, W. Fundamenski, A.C.C. Sips and EFDA-JET Contributors, \textit{Nucl. Fusion} \textbf{52}, 103016 (2012).

\bibitem{Karney_1986}
C.F.F. Karney, \textit{Comput. Phys. Rep.} \textbf{4}, 183 (1986).

\bibitem{Lloyd_1996}
B. Lloyd, P.G. Carolan and C.D. Warrick, \textit{Plasma Phys. Control. Fusion} \textbf{38}, 1627 (1996).

\bibitem{Wesson_2004}
J. Wesson, \textit{Tokamaks} (Oxford: Clarendon Press, 2004).

\bibitem{OBrien_1986}
M.R. O'Brien, M. Cox and D.F.H. Start, \textit{Nucl. Fusion} \textbf{26}, 1625 (1986).

\bibitem{Shevchenko_2015b}
V.F. Shevchenko \textit{et al.}, arXiv:1501.01798 [physics.plasm-ph] (2015).

\bibitem{Urban_2011}
J. Urban \textit{et al.}, \textit{Nucl. Fusion} \textbf{51}, 083050 (2011).

\bibitem{Wauters_2011}
T, Wauters \textit{et al.}, \textit{Plasma Phys. Control. Fusion} \textbf{53}, 125003 (2011).

\bibitem{Wong_1980}
K.L. Wong, R. Horton and M. Ono, \textit{Phys. Rev. Lett.} \textbf{45}, 117 (1980).















\end{thebibliography}
\end{document}